\documentclass[notitlepage,amsmath,amssymb,pra,twocolumn,floats,aps,superscriptaddress]{revtex4-2}

\usepackage{latexsym, amssymb, amsmath, titlesec, mathtools, hyperref, braket, accents, xspace, setspace, graphicx, natbib, epsfig, url, mathrsfs, longtable, xcolor}
\usepackage[normalem]{ulem}
\usepackage{cancel}

\usepackage{comment}

\newcommand{\arctanh}{\text{arctanh}}

\newcommand{\toto}[2]{\underset{\overset{#1\to #2}{}}{\longrightarrow }}
\newcommand{\dif}{\text{d}}
\newcommand{\fdif}{\text{D}}
\newcommand{\e}{\text{e}}
\newcommand{\im}{\text{i}}

\newcommand{\R}{\mathbb{R}}

\newcommand{\fL}{\mathcal{L}}

\newcommand{\q}{\mathrm{q}}

\newcommand{\tp}{\mathrm{p}}
\newcommand{\td}{\mathrm{d}}

\newcommand{\tl}{\mathrm{l}}
\newcommand{\tg}{\mathrm{g}}
\newcommand{\tQ}{\mathrm{Q}}
\newcommand{\tP}{\mathrm{P}}
\newcommand{\tS}{\mathrm{S}}
\newcommand{\tB}{\mathrm{B}}

\newcommand{\tD}{\mathrm{D}}
\newcommand{\tF}{\mathrm{F}}

\newlength{\dhatheight}
\newcommand{\doublehat}[1]{%
    \settoheight{\dhatheight}{\ensuremath{\hat{#1}}}%
    \addtolength{\dhatheight}{-0.35ex}%
    \hat{\vphantom{\rule{1pt}{\dhatheight}}%
    \smash{\hat{#1}}}}

\newcommand{\inst}{\mathrm{inst}}
\newcommand{\tmin}{\mathrm{min}}

\def\0{^{(0)}}
\def\1{^{(1)}}
\def\2{^{(2)}}

\def\cH{\mathcal{H}}

\def\cD{\mathcal{D}}

\usepackage{braket}

\def\st{_\mathrm{st}}

\definecolor{DarkGreen}{rgb}{0.0,0.5,0.0}

\begin{document}
\title{Spectroscopy of Quantum Phase Slips: Visualizing Complex Real-Time Instantons}
\author{Foster Thompson}
\affiliation{School of Physics and Astronomy, University of Minnesota, Minneapolis, MN 55455, USA}
\affiliation{Institute for Theoretical Physics, University of Cologne, 50937 Cologne, Germany}
\author{Daniel K. J. Bone\ss}
\affiliation{Department of Physics, University of Konstanz, 78464 Konstanz, Germany}
\author{Mark Dykman}
\affiliation{Department of Physics and Astronomy, Michigan State University, East Lansing, MI 48824, USA}
\author{Alex Kamenev}
\affiliation{School of Physics and Astronomy, University of Minnesota, Minneapolis, MN 55455, USA}
\affiliation{William I. Fine Theoretical Physics Institute, University of Minnesota, Minneapolis, MN 55414, USA}

\begin{abstract}
Parametrically driven oscillators can emerge as a basis for the next generation of qubits. Classically, these systems exhibit two stable oscillatory states with opposite phases. Upon quantization, these states turn into a pair of closely spaced Floquet states, which can serve as the logical basis for a qubit. However, interaction with the environment induces phase-slip events which set a  limit on qubit coherence.  Such phase slips persist even at zero temperature due to a mechanism known as quantum activation \cite{QuantumActivation}. In contrast to conventional tunneling, the quantum activation is described by a {\em real-time}  instanton trajectory in the complexified phase space of the system.
In this work, we show that the  phase-slip rate is exponentially sensitive to weak AC perturbations. The spectrum  of the system's response -- captured by the so-called logarithmic susceptibility (LS) -- enables a direct observation of characteristic features of  real-time instantons.  Studying this spectrum suggests new means of efficient qubit control.
\end{abstract}
\maketitle

\section{Introduction}

Studies of  parametrically driven oscillators (PDO) have been gaining significant attention in the field of quantum information processing \cite{Proposal_Encode_Qubits, CatQubitBackground2, ExprQPDO2, CatQubitBackground3, Adiabatic_quantum_computing_nonlinear_network, Preparation_of_states_PDO, Driving_into_chaos_regime_validity,Exp_cancellation_tunneling,Exp_large_bit_flip,Exp_two_photon_dissipation}. These systems provide novel methods for metrology \cite{Sensing_classical_1, Sensing_classical_2, Sensing_classical_3,Eichler2023, Quantum_metrology} and annealing \cite{Annealing_1, Annealing_2, Annealing_3, Annealing_4, Annealing_5, Annealing_6}, both in the classical and quantum domain.  The main interest, however,  stems from their ability to serve as a robust platform for encoding qubits through their non-equilibrium steady states \cite{Proposal_Encode_Qubits, 2PhotonLossExp3, ExprQPDO2, ExprQPDO3, CatQubitBackground2,CatQubitBackground3}. Classically,  oscillators parametrically driven at frequency $2\omega_\tp$, exhibit {\em two} steady  states of forced vibrations with equal amplitudes and  opposite phases \cite{Landau_book_1}. In the quantum regime, these two states correspond to generalized coherent states with the opposite sign of their complex amplitudes \cite{CatQubitBackground3}. Their symmetric and antisymmetric superpositions form the Floquet eigenstates, which may be used as a cat qubit.
Both classical and quantum versions of this system have been extensively studied \cite{ClassicalActivation,ClassicalLS,QuantumActivation,QuantumSymmetryLifting,QuantumSymmetryLiftingZeroT,0TKerrOscillators}.
They have also been realized in a variety of experiments on mesoscopic systems, 
for example in electrons in Penning traps \cite{Penning1,Penning2} or more recently in the context of bosonic qubits constructed from superconducting circuits and resonators \cite{ExprQPDO1,ExprQPDO2,Exp_large_bit_flip,ExprQPDO3}, see the book \cite{Eichler2023} for more references.



Fluctuations, either thermal or quantum, induce transitions between the two dynamically stable states. Since such transitions lead to a sudden change of the oscillation phase by $\pi$, they are the phase slips of the steady state oscillations. These events provide the fundamental limit for the qubits' coherence time. It is therefore extremely important to understand a rate of their occurrence to find ways to reduce it.  
The corresponding phase-slip  rates for both  classical \cite{ClassicalActivation} and quantum \cite{QuantumActivation} PDO can be evaluated with the help of an instanton-based calculation using a path integral description.  Such instantons evolve in real time as opposed to imaginary-time tunneling events in equilibrium quantum mechanics. They are reminiscent of the most probable trajectories followed by a Brownian particle as it escapes from a potential well \cite{Kramers1940}. The important distinction, though, is that a PDO is a nonequilibrium system with its  stable states being periodic oscillations, and the noise can be of quantum origin. The resulting rates are exponentially sensitive to various system parameters, such as detuning of half the modulation frequency $\omega_p$ from the oscillator eigenfrequency, $\omega_0$, the oscillator nonlinearity, dissipation rate, etc. 

It is therefore important to gain a maximally detailed information about phase slips internal dynamics to minimize the rate of their occurrence.  This in turn  maximizes the coherence time of qubits based on PDOs. It is also a major challenge to find a way to ``visualize'' a quantum  real-time instanton, which would provide a direct proof of the mechanism of quantum activation.

A way of doing it is to use spectroscopy. In the present context, it amounts to applying a weak external drive (in addition to the relatively strong parametric drive) with some frequency $\omega_\td$, and monitoring how the phase-slip rate is affected by such extra signal. 
The dependence of the phase slip rate on the spectroscopic frequency, $\omega_\td$, reflects the coherent internal dynamics of the individual phase slips. We are not aware of any other means of accessing this information, given that the real-time instanton is a trajectory in the complex phase space of a driven oscillator, cf. \cite{Dykman1988a}.   

\begin{figure}[t]
    \centering
    \scalebox{1}{\includegraphics{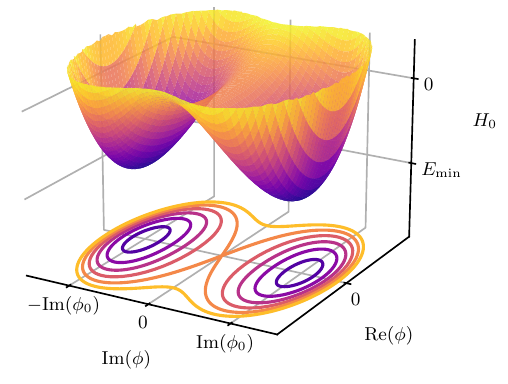}}
    \caption{Rotating wave quasi-potential $E=H_0(\bar\phi,\phi)$ of the PDO. The two minima correspond to the two out-of-phase period-two oscillations.}
    \label{fig:H0Surface}
\end{figure}

Qualitatively, the phase-slips can be understood in the following way. In the frame rotating at twice the modulation period $2\pi/\omega_p$, the unperturbed PDO may be described as a quantum dissipative oscillator in a stationary quasi-potential depicted in Fig.~\ref{fig:H0Surface}. The coordinates are given by components of the complex coherent state, $\phi$, which serve as a 2D phase-space of the corresponding dynamical system. For a sufficiently strong parametric driving and a weak damping, phase-slips proceed as slowly expanding spirals rotating around one of the two minima with frequency $\omega(E)$ (here $E$ is the quasi-energy, depicted in Fig.~\ref{fig:H0Surface}). The corresponding rotation frequency, $\omega(E)$, is maximal near the bottom of the well, where it takes the value $\omega_\tmin=\omega(E_\tmin)$. It monotonically decreases with increasing values of $E$ and goes to zero when the trajectory approaches the saddle point between the two wells. 

When the weak spectroscopic drive is in resonance with the period-two parametric oscillations, $\omega_\td=\omega_\tp$, the rotating frame motion is perturbed by a {\em time-independent} tilting of the double-well quasi-potential. Such a perturbation breaks the symmetry between the two oscillatory states and makes switching out of the shallow well exponentially more frequent. The corresponding change in the (logarithm of the) phase-slip rate was obtained for both classical \cite{ClassicalLS} and quantum \cite{QuantumSymmetryLifting,QuantumSymmetryLiftingZeroT} PDO models. When the spectroscopic drive is off-resonance with the parametric oscillations, the rotating wave perturbation is time-dependent with an effective frequency $\nu=\omega_\td-\omega_\tp$. The latter may resonate with the internal rotating motion at some quasi-energy along the phase-slip trajectory, $\nu=\omega(E)$. This in turn leads to a strong enhancement of the switching rate that may be significantly larger than the corresponding $\nu=0$ effect.


\begin{figure}[ht]
    \centering
    \includegraphics[width=1\linewidth]{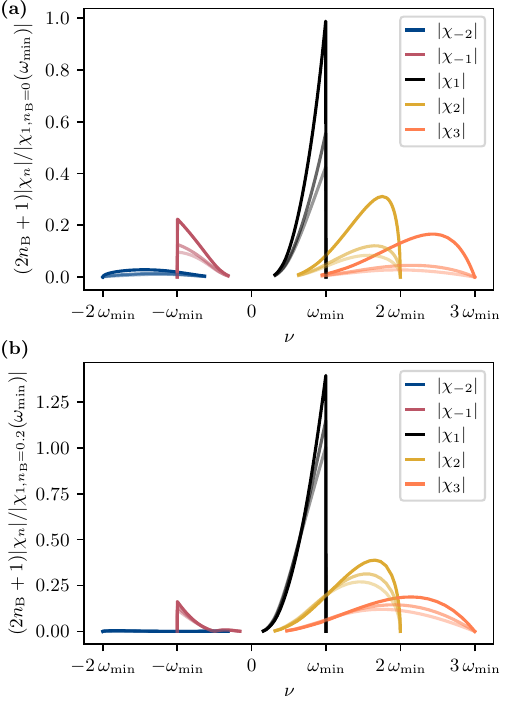}
    \caption{Examples of the quantum LS shown for two different values of detuning, $\Delta\propto \omega_0-\omega_p$, scaled by the modulation amplitude $\propto \lambda$; fainter lines correspond to smaller temperatures;  $n_\tB$ is the thermal occupation number of the oscillator in the absence of modulation.  Panel (a) shows $\Delta/2\lambda=0.6$ for $T=0$, $n_\tB=0.5$, and the classical limit.  Panel (b) shows $\Delta/2\lambda=-0.6$ for $n_\tB=0.2$, $n_\tB=0.5$, and the classical limit.  The $T=0$ result from the top plot is obtained using the $T=0$ instanton from Eq.~(\ref{pinstT=0}), while all finite temperature results are obtained by solving Eq.~(\ref{K0quantum}) numerically for $K(I,p)=0$.  The classical limits show Eq.~(\ref{LSharmonicClassical}) rescaled by $2T/\omega_\tp$.
    In the bottom plot, the vanishing of $\chi_n$ for $n<0$ at frequencies larger than $n\omega_\tmin$ can be observed for all temperatures; this is discussed in detail below, see Fig. \ref{fig:Classical_Log_Susc}.}
    \label{fig:QLS}
\end{figure}


In this work, we address spectroscopy of quantum phase-slips in a broad range of environmental temperature, from the temperatures where the thermal occupation number $n_\tB$ of the oscillator in the absence of modulation is zero to the classical high-temperature limit, $n_\tB\gg 1$.  We show that the strong enhancement of the switching rate persists for all temperatures. It exhibits a strongly pronounced frequency dependence, which  serves as an experimentally detectable signature of the quantum phase-slips' coherent dynamics.

We show that the resonant enhancement of the phase-slip rate occurs only when the spectroscopic frequency in the rotating frame is inside the frequency band of the intrawell motion, $|\nu| < \omega(E_\tmin)$, i.e. when there exists a value of the quasi-energy inside the well for which the condition $\nu=\omega(E)$ is met. There is a sharp (non-analytic as a function of $\nu$) drop of the response when $|\nu|$ exceeds the largest intrawell frequency, $\omega_\tmin=\omega(E_\tmin)$,  corresponding to the disappearance of the resonance. However, there are also secondary sharp features in the response at $\nu = n\omega_\tmin$, where $n$ is an integer, corresponding to resonances with higher harmonics of the oscillatory instanton trajectory, see Fig.~\ref{fig:QLS}. Such higher harmonics exist due (and are highly sensitive) to the anharmonicity of the quasi-potential, Fig.~\ref{fig:H0Surface}.  
The corresponding classical (i.e. for thermal phase-slips) calculation was presented in Ref.~\cite{ClassicalLS}.

The technical approach we take is to develop a parallel treatment of the stochastic classical and quantum PDO based on the path integral representation of the Keldysh technique \cite{KamenevBook}. This has an advantage of treating both the classical and quantum theories on the same footing.  Moreover, the classical theory constitutes a particular limiting case at the level of the quantum action. This observation provides a transparent connection between the classical and quantum regimes before any specific calculations are performed. Corrections to the switching rate {\em exponent} are then straightforward to obtain for both the classical and quantum theory via the computation of a certain response function known as the logarithmic susceptibility \cite{LSoriginal}.

The microscopic models for both the quantum and classical PDO are introduced in Section \ref{II} and the corresponding path integral treatments are presented. Section \ref{III} discusses the computation of the bare switching rate using action-angle coordinates.  Previously known results for both  classical \cite{ClassicalActivation} and quantum \cite{QuantumActivation} problems are re-derived here for completeness.   The logarithmic susceptibility is discussed in section \ref{IV}, for which the classical result \cite{ClassicalLS} is again re-derived and the quantum result is newly obtained. Section \ref{V} discusses the near vicinity of the PDO instability (bifurcation) point, where the wells are shallow. In this regime the dissipation necessarily becomes fast compared to the coherent rotation and the motion becomes overdamped, resulting in the breakdown of the resonant response. We derive the spectroscopic signatures of this regime. Concluding remarks are given in Section \ref{VI}.

\section{The Parametric Oscillator}\label{II}
We consider a single  mode with a frequency $\omega_0$ and the quartic Duffing anharmonicity $G$ subject simultaneously to a parametric drive with frequency $2\omega_\tp$ and a weak spectroscopic signal at a different frequency, $\omega_\td$ (see Appendix~\ref{sec:table} for the relation to the notations in \cite{QuantumActivation,QuantumSymmetryLifting,QuantumSymmetryLiftingZeroT}). The classical Hamiltonian describing this system is given by
\begin{multline}\label{HamPDO}
H=\frac{1}{2}\,p^2+\frac{1}{2}\big(\omega_0^2+A_\tp\cos(2\omega_\tp t)\big)x^2+\frac{1}{4}Gx^4\\+A_\td\cos(\omega_\td t+\varphi_\td)x,
\end{multline}
where the mass of the oscillator has been absorbed into a rescaling of the momentum and thus set to unity for brevity. 

When the amplitude, $A_\td$, of the second drive is small, it may be treated as a perturbation. The dynamics can be well-approximated using the rotating wave approximation in the frame of the parametric drive. This is conveniently achieved using the complex coordinate in the rotating frame 
\begin{equation}
    \phi=\sqrt{\frac{\omega_\tp}{2}}\,(x+\im p/\omega_\tp)\,\e^{-\im\omega_\tp t}.
\end{equation}
Upon dropping the fast counter-rotating terms, the Hamiltonian in the rotating frame $H_\mathrm{RWA}=H_0+H_1$ becomes 
\begin{subequations}\label{HRWA}
\begin{equation}\label{H0}
H_0(\bar\phi,\phi)=\Delta\bar\phi\phi+\lambda(\phi^2+\bar\phi^2)+\frac{g}{4}(\bar\phi\phi)^2;
\end{equation}
\begin{equation}
H_1(\bar\phi,\phi;t)=\alpha(\e^{\im(\nu t+\varphi_\td)}\phi+\e^{-\im(\nu t+\varphi_\td)}\bar\phi),
\end{equation}
\end{subequations}
where $\alpha=A_\td/(2\sqrt{2\omega_\tp})$ is the normalized amplitude of the (weak) spectroscopic signal, $\Delta=(\omega_0^2-\omega_\tp^2)/2\omega_\tp$ is a detuning,  $\lambda=A_\tp/8\omega_\tp$ is the normalized amplitude of the (strong) parametric drive, $g=3G/2\omega_\tp^2$ is an anharmonicity, and $\nu=\omega_\td-\omega_\tp$ is the effective spectroscopic frequency. The complex variables $\phi$ and $\bar\phi$ form a canonical pair with  the corresponding equations of motion  $i\partial_t\phi = \partial_{\bar\phi} H_\mathrm{RWA}$ and $i\partial_t{\bar\phi} = -\partial_{\phi} H_\mathrm{RWA}$.

The classical dynamics described by the unperturbed rotating-wave approximated Hamiltonian $H_0$ is symmetric under the  parity symmetry $\phi\to-\phi$. For a fixed quasi-energy, $E=H_0(\bar\phi,\phi)$, in the laboratory frame the oscillator may have period-two states, which oscillate at frequency $\omega_\tp$, modulated by $\omega(E)$.  The nature of the state with the minimal quasi-energy depends on the dimensionless ratio $\Delta/2\lambda$.  For $\Delta/2\lambda>1$ there is a unique symmetric minimum of $H_0$ at $\phi=0$, corresponding to the absence of the parametrically induced oscillations of the mode. For $\Delta/2\lambda<1$ the $\phi=0$ point becomes a saddle point and $H_0$ acquires a double well shape\footnote{
More precisely, we consider $|\Delta|<2\lambda$. If $\Delta$ is too strongly negative, $\Delta<-2\lambda$, the point at $\phi=0$ becomes a maximum of $H_0$ and two additional saddles appear at $\phi=\pm\sqrt{-2(2\lambda+\Delta)/g}$. We do not consider the large negative detuning regime in this work.}, depicted in Fig.~\ref{fig:H0Surface}.
The parity symmetry is spontaneously broken and the system displays bistability as there emerge two oscillatory states, which are stable in the presence of dissipation.
They are located at the minima of the function $H_0$, at $\phi=\pm\phi_0$ with $\phi_0=\im\sqrt{2(2\lambda-\Delta)/g}$ and $H(\pm\phi_0)\equiv E_\tmin=-(2\lambda-\Delta)^2/g<0$. Near the bottom of the wells, the steady states $\omega_\tp$-oscillations are modulated by small-amplitude vibrations with the frequency
\begin{equation}
    \omega_\tmin=\omega(E_\tmin)=2\sqrt{2\lambda(2\lambda-\Delta)}, 
\end{equation}
which play an important role in the PDO spectroscopy.

The nature of the classical motion described by $H_0$ away from the fixed points depends on the value of the quasi-energy.
For $E\in[E_\tmin,0)$ all classical states are localized in one of the two wells.  For positive quasi-energies, $E>0$, all states regain their symmetry and are not restricted to a single well. At $E=0$ there is a separatrix separating these two regimes, which passes through the saddle point at $\phi=0$. For our purposes, it suffices to consider in detail only the intrawell motion with $E\in[E_\tmin,0)$. Solutions of the classical Hamilton's equations $\im\partial_t\phi=\partial_{\bar\phi}H_0$, centered around $\phi_0$, are denoted with an underline, $\underline\phi(t;E)$.
These solutions are elliptic functions; details of their specific functional forms are presented in Appendix \ref{appA}. As elliptic functions, the solutions $\underline\phi(t;E)$ are bi-periodic in complex plane of the time argument $t$.  For all values of the system parameters and quasi-energy, there is a real-valued period $t_1(E)$ and an additional complex period $t_2(E)$, so that
\begin{equation}
\underline\phi\big(t+t_j(E);E\big)=\underline\phi(t;E)
\end{equation}
for $j=1,2$. The real period is the physical period of the motion in the rotating frame, $t_1(E)=2\pi/\omega(E)$.  The complex period does not have a meaning in the classical problem, however its appearance can be traced to the fact that constant quasi-energy condition $H_0(\bar\phi,\phi)=E$ defines an elliptic curve in the complexified classical phase space.  As a consequence of this fact, the complex period is related to the tunneling rate of a closed quantum system with the Hamiltonian $H_0$, for example see Refs.~\cite{GuldenTori1,GuldenTori2}.

The second drive, $H_1$, explicitly breaks the parity symmetry.  
In the bistable regime this results in the modulation of the heights of the two wells. When the second drive period is exactly at half-resonance with the parametric drive, $\nu=\omega_\td-\omega_\tp=0$, the problem is time-independent and the two wells are tilted so that one is shallower than the other. The direction of this tilt is dictated by the phase $\varphi_\td$ between the parametric drive and  the second drive.
Off resonance, the heights of the two wells oscillate in time, so that a given well is tilted to become shallower and then deeper than its counterpart over the oscillation period $2\pi/\nu$.

\subsection{Action-Angle Variables}

As will be discussed in detail in section \ref{III}, the problem of computing the quantum fluctuation-induced switching rate involves identifying appropriate instanton trajectories that connect the classical fixed points.  This is in general a difficult task for driven systems, because the rotating wave Hamiltonian, $H_0$, cannot be written as a sum of kinetic and potential energies.  Thus the momentum coordinate cannot be integrated out, and so the resulting effective semiclassical phase space has large dimension.
This predicament can be avoided by  identifying  a coordinate system which separates degrees of freedom that evolve with different time scales.  When this is possible, the faster coordinate can be effectively integrated out to provide an effective dynamics of the slow variable, thus reducing the dimensionality of the phase space. This greatly simplifies the computational challenge associated with finding the instanton trajectory.

In the present context, 
the action-angle canonical coordinates, $(\phi,\bar\phi) \to (I,\theta)$, provide such a decomposition.  They are defined in a way to make the unperturbed Hamiltonian angle independent, 
$H_0(\bar\phi,\phi)=H_0(I)=E(I)$, and preserve the canonical Poisson bracket between $I$ and $\theta$.  The corresponding Hamilton equation
\begin{equation}
\partial_t\theta=\partial_I H_0(I)\equiv \omega(I)
\end{equation}
dictates $\theta=\omega(I)t$, while $I=I(E)$ is an integral of motion for $H_0$. The action for a given quasi-energy $E$ is given by the area enclosed by the trajectory $\underline\phi(t;E)$ in the complex $\phi$-plane  
\begin{equation}
\label{eq:action} 
I(E) =\int\limits_0^{t_1(E)}\frac{\dif t}{2\pi}\ \underline{\bar\phi}(t;E)\, \im\partial_t\underline\phi(t;E).
\end{equation}
The function $I(E)$ is thus monotonic with $I(E_\tmin)=0$ and $I(E=0)=I_\mathrm{top}>0$, cf. Fig.~\ref{fig:H0Surface}. 
The frequency $\omega(I)$ is also a  monotonic function, decreasing from $\omega(0)\equiv\omega_\tmin>0$ down to $\omega(I_\mathrm{top})=0$. The  corresponding periodic classical trajectory may be also transformed to the action-angle representation and decomposed into the Fourier series:   
\begin{equation}\label{ActionAngle}
\underline\phi(t;E)=\underline\phi(\theta(t);I)=\sum_nc_n(I)\, \e^{-\im n\theta(t)}.
\end{equation}
The functional form of the Fourier coefficients, $c_n(I)$, and their asymptotic expressions are given in Appendix \ref{appA}.

\subsection{Classical Stochastic Theory}

The influence of an environment can be modeled by a coupling between the system and a thermal bath of linear oscillators with position coordinates $X_u$ which oscillate with a spectrum of frequencies $\omega_u$.  Each bath mode is taken to be in thermal equilibrium at temperature $T$ (here and throughout we work with units in which $k_\tB=1$).  We consider a linear coupling between the system and the environment described by the interaction $H_\mathrm{int}=\sum_uc_uX_ux$.  This retains the parity symmetry of the un-perturbed system $x\to-x$ if it is expanded also to include the bath modes $X_u\to-X_u$.

In the standard way \cite{ClassicalLangevin1}, averaging over the bath degrees of freedom gives an effective theory described by the stochastic Langevin equation in the rotating frame and in the rotating wave approximation
with dissipation and noise. The dissipation strength is given by $\kappa=J(\omega_\tp)/2$ where $J(\epsilon)=\pi\sum_u(c_u^2/\epsilon)\delta(\epsilon-\omega_u)$ is the bath spectral density. The Langevin equation reads 
\begin{equation}\label{LangevinPDO}
\im\partial_t\phi=\partial_{\bar\phi}H_\mathrm{RWA}-\im\kappa\phi+\xi(t),
\end{equation}
where the Hamiltonian, $H_\mathrm{RWA}$,   is given by Eq.~(\ref{HRWA}) and $\xi(t)$ is a complex Gaussian white noise with zero mean and equal variance for real and imaginary parts $\braket{\bar\xi(t)\xi(t')}=(2\kappa T/\omega_\tp)\delta(t-t')$ and $\braket{\xi(t)\xi(t')}=0$.

With this, one may pass to the action-angle variables by taking the time derivative of $I=I(E(\underline{\bar\phi},\underline\phi))$ and then using Eq.~(\ref{LangevinPDO}).
This results in a Langevin equation for the action variable with a multiplicative noise,
\begin{equation}\label{LangevinI}
\partial_tI=\mathcal{E}(I,\theta)+\bar\eta(I,\theta)\xi(t)+\eta(I,\theta)\bar\xi(t),
\end{equation}
where the functions $\mathcal{E}=\mathcal{E}_0+\alpha\mathcal{E}_1$ and $\eta$ are given by
\begin{subequations}\label{LangevinIFncs}
\begin{equation}
\mathcal{E}_0(I,\theta)=-\frac{\kappa}{\omega(I)}\,\big(\underline\phi\,\partial_\phi H_0+\underline{\bar\phi}\,\partial_{\bar\phi}H_0\big),
\end{equation}
\begin{equation}
\mathcal{E}_1(I,\theta;t)=\partial_\theta\underline\phi\,\partial_\phi H_1+\partial_\theta\underline{\bar\phi}\,\partial_{\bar\phi}H_1,
\end{equation}
\begin{equation}
\eta(I,\theta)=\frac{\im\partial_{\bar\phi}H_0}{\omega(I)}.
\end{equation}
\end{subequations}
The angle coordinate can be treated similarly.  In the limit of weak dissipation $\kappa\ll\omega_\tmin$, the motion is underdamped.  Therefore the action coordinate undergoes slow relaxation with a time scale set by the weak dissipation, $\kappa$, while the angle coordinate coherently rotates at a faster scale, $\omega(I)$, set by the dissipationless classical problem.  The dynamics of the phase can thus be well-approximated by neglecting the effects of dissipation and fluctuations,
\begin{equation}\label{LangevinTheta}
\partial_t\theta\simeq\omega(I).
\end{equation}
This separation of time scales means the dynamics of the action can be described by averaging over the fast angle rotation. 

We are interested in rare events of phase-slip transitions, i.e., switching between the classically stable states with opposite phase. To calculate the phase-slip rate $W_\mathrm{ps}$ it is convenient to first pass to the path integral description and then use the WKB approximation. We do it here in the spirit of the Keldysh formalism \cite{KamenevBook,SDKeldyshLindblad,FieldTheoryLindblad}. We introduce the partition function as noise average of equations Eqs.~(\ref{LangevinI}) and (\ref{LangevinTheta}), enforced through delta functions:
\begin{align}\label{ZItheta}
Z=&\int\fdif\bar\xi\fdif\xi\,\exp\left[-\frac{\omega_\tp}{2\kappa T}\int\dif t\bar\xi\xi\right]\nonumber\\
&\times\int\fdif I\fdif\theta\,\delta\big(\partial_tI-\mathcal{E}-\bar\xi\eta-\bar\eta\xi\big)\delta\big(\partial_t\theta-\omega\big).
\end{align}
This can be massaged into the standard form by introducing a Lagrange multiplier field, which will be denoted $p$, to impose the delta constraint of the action coordinate, and then integrating out both the noise and the angle.  Integrating out the angle leads to the replacement $\theta\to\underline\theta(I)=\int^t\dif t\,\omega(I)$, thus resulting in an effective theory with only two fields:
\begin{subequations}\label{IpTheory}
\begin{equation}
Z=\int\fdif I\,\fdif p\,\, \e^{\im S[I,p]}
\end{equation}
\begin{equation}
\im S[I,p]=-\int\dif t\big(p\partial_tI-K(I,p)\big),
\end{equation}
\end{subequations}
where the standard integration contour of the auxiliary variable runs along the imaginary axis, $p\in\im\R$.

The effective Hamiltonian, $K=K_0+K_1$, is a sum of the  term $K_0$ that describes dissipation  and the term $K_1$, which is due to the second drive. The term $K_0$ is obtained by averaging over the fast-rotating parts of the functions from Eq.~(\ref{LangevinIFncs}).
This is achieved using the Fourier series representation of the classical solutions Eq.~(\ref{ActionAngle}) and then using the replacement $\exp(\im(n'-n)\underline\theta(I))\to\delta_{nn'}$ to re-sum the resulting series.  This results in the substitutions $\mathcal{E}_0\to-2\kappa I$ and $\bar\eta\eta\to\Gamma(I)$, where $\Gamma(I)=\int_0^{2\pi}(\dif\theta/2\pi)\, \partial_\theta\underline{\bar\phi}(\theta;I)\,\partial_\theta\underline\phi(\theta;I)$ is a function whose properties may be obtained in certain limits, see Appendix \ref{appA}. Employing Eqs.~(\ref{eq:action}) and (\ref{ActionAngle}), one observes the following expressions through the Fourier coefficients:
\begin{equation}
\label{eq:sum-rules}
    I=\sum_n n|c_n(I)|^2; \qquad
    \Gamma(I)=\sum_n n^2 |c_n(I)|^2
\end{equation}
The resulting unperturbed effective Hamiltonian is  given by
\begin{align}\begin{split}\label{K0classical}
K_0(I,p)&=p\,\mathcal{E}_0(I)+\frac{2\kappa T}{\omega_\tp}\, \bar\eta(I)\eta(I)\,p^2\\&\simeq-2\kappa p I+\frac{2\kappa T}{\omega_\tp}\,\Gamma(I)\,p^2.
\end{split}\end{align}
This is the effective Hamiltonian for a Langevin equation describing the linear relaxation of the action variable, $\partial_t I=-2\kappa I$, under the influence of a multiplicative noise with the variance $\sim(\kappa T/\omega_\tp)\Gamma(I)$.

The second-drive induced term $K_1$ does not have any fast rotating terms that can be neglected, so all Fourier modes should be kept.  It has the form:
\begin{equation}\label{K1classical}
K_1(I,p;t)=\alpha p\sum_nnc_n(I)\e^{-\im(n\underline\theta(I)-\nu t-\varphi_\td)}+\mathrm{c.c.}
\end{equation}
where $\mathrm{c.c.}$ is complex conjugate.

\subsection{Quantum Theory}

The Hamiltonian operator $\hat H$ of the quantum theory is defined by adorning $x$ and $p$ in Eq.~(\ref{HamPDO}) with hats.
The quantum Hamiltonian in the rotating wave approximation is similarly obtained from Eq.~(\ref{HRWA}) by replacing the complex coordinates $\phi$ and $\bar\phi$ with the bosonic creation and annihilation operators $\hat a$ and $\hat a^\dagger$, which satisfy the canonical bosonic commutation rule $[\hat a,\hat a^\dagger]=1$.

Like in the classical theory, the environment may be modeled by a thermal bath of linear oscillators linearly coupled to the system.
Integrating out the bath modes gives an effective dynamics of the reduced density matrix of the system described, in the rotating wave approximation, by a Lindblad equation \cite{KamenevBook},
\begin{align}
\label{Lindbladian}
\partial_t\rho=&\doublehat\fL\rho=-\im[\hat H_\mathrm{RWA},\rho]\nonumber\\
&+\sum_{v=\ell,\tg}\gamma_v\Bigl(\hat L_v\rho\hat L_v^\dagger-\frac{1}{2}\{\hat L_v^\dagger\hat L_v,\rho\}\Bigl),
\end{align}
where the Lindbladian superoperator $\doublehat\fL$ acts on the Hilbert space of operators.  There are two jump operators $\hat L_\ell=\hat a$ and $\hat L_\tg=\hat a^\dagger$ with the corresponding rates $\gamma_\ell=2(n_\tB+1)\kappa$ and $\gamma_\tg=2n_\tB \kappa$.
The Bose occupation number $n_\tB=1/(\e^{\omega_\tp/T}-1)$ is the occupation number of the bath in the rotating frame given by the Bose function evaluated at $\epsilon=\omega_\tp$. The coupling leads also to renormalization of the oscillator eigenfrequency, which we assume to have been incorporated into $\omega_0$.

The path integral description for the quantum problem is also provided by the Keldysh formalism \cite{KamenevBook,SDKeldyshLindblad,FieldTheoryLindblad}.  The partition function for the system is represented as a coherent state functional integral
\begin{equation}  Z=\int\fdif\phi^\pm\fdif\bar\phi^\pm\exp(\im S[\bar\phi^\pm,\phi^\pm]),  
\end{equation} 
where the action corresponding to the Lindbladian from Eq.~(\ref{Lindbladian}) is
\begin{multline}
\label{eq_ActionS_full}
S[\bar\phi^\pm,\phi^\pm]=\int\dif t\Big(\bar\phi^+ \im\partial_t\phi^+-\bar\phi^-\im\partial_t\phi^-\\-H_\mathrm{RWA}^++H_\mathrm{RWA}^--\im\sum_{v=\ell,\tg}\gamma_vD_v\Big),
\end{multline}
where $H_\mathrm{RWA}^\pm=H_\mathrm{RWA}(\bar\phi^\pm,\phi^\pm)$ and the dissipative terms are
\begin{subequations}
\begin{equation}
D_\ell(\bar\phi^\pm,\phi^\pm)=\bar\phi^-\phi^+-\frac{1}{2}(\bar\phi^+\phi^++\bar\phi^-\phi^-),
\end{equation}
\begin{equation}
D_\tg(\bar\phi^\pm,\phi^\pm)=\bar\phi^+\phi^--\frac{1}{2}(\bar\phi^+\phi^++\bar\phi^-\phi^-).
\end{equation}
\end{subequations}

The action-angle coordinates may be introduced in the Keldysh language using the coordinate transformation defined by Eq.~(\ref{ActionAngle}) applied for the fields on forward and backward parts of the time contour independently.  The resulting set of four coordinates $I^\pm$ and $\theta^\pm$ can be separated into physical and auxiliary ``quantum" variables with the Keldysh rotation: $I^\pm=I\pm I^\q/2$ and $\theta^\pm=\theta\pm\theta^\q/2$.  In total, this amounts to the coordinate transformation:
\begin{equation}\label{ActionAngleKeldysh}
\phi^\pm=\sum_nc_n(I\pm I^\q/2)\e^{-\im n(\theta\pm\theta^\q/2)},
\end{equation}
where, as before, $c_n(I)$ are Fourier coefficients of the classical trajectory with the fixed action, $I$. 

Like in the classical theory, for weak dissipation the fluctuations of the angle coordinate can be neglected.  This is achieved by expanding to first order in the quantum action coordinate $I^\q$ in the Hamiltonian, so that $H_0^+-H_0^-\simeq\partial_IH_0(I)I^\q=\omega(I)I^\q$, and dropping the $I^\q$ dependence in the perturbation $H_1^\pm$ and dissipative parts $D_v$.  With this, $I^\q$ appears only at linear order in the action $\sim I^\q(\partial_t\theta-\omega(I))$.  Subsequently integrating out both $I^\q$ and $\theta$ thus leads to the replacement $\theta\to\underline\theta(I)=\int^t \!\dif t \,\omega(I)$ like in the classical theory.
This defines an effective theory with only two fields, which upon defining $\theta^\q=-\im p$ has the same form as Eq.~(\ref{IpTheory}).

The effective Hamiltonian $K=K_0+K_1$ comes from the dissipative part of the Lindbladian evolution $D_v$ and the second drive perturbation $H_1$.
Averaging over the fast-rotating parts of $K_0$ is carried out in the same way as in the classical theory.  It results in the replacements $\bar\phi^\mp\phi^\pm\to\sum_n|c_n|^2\e^{\mp np}$ and $\bar\phi^\pm\phi^\pm\to\sum_n|c_n|^2$.  The result is an effective Hamiltonian with a form matching that of reaction kinetics models
\begin{equation}\label{K0quantum}
K_0(I,p)=\sum_nW_n(I)(\e^{-np}-1).
\end{equation}
This describes the stochastic dynamics of the action variable in which each term in the sum with $n\gtrless0$ is associated with a processes in which the quasi-energy transitions up or down by $n$ quantized levels.  The transitions are induced by the bath and their rates are given by:
\begin{equation}\label{Wn}
W_n(I)=\gamma_\ell|c_n(I)|^2+\gamma_\tg|c_{-n}(I)|^2.
\end{equation}
There are transitions both down and {\em up} in quasi-energy even for $T=0$ (which corresponds to $\gamma_\tg=0$). The $T=0$ up-transitions are due to the parametric driving, which makes the quasi-energy states to be different from the energy states of the corresponding equilibrium oscillator.  This is the formal manifestation of the phenomenon known as quantum heating, in which driven-dissipative quantum systems can have finite occupation even when connected to a zero temperature bath \cite{QuantumActivation}.

In this formulation, it is very clear how the classical theory can be obtained as a limit from the quantum theory.  Expanding Eq.~(\ref{K0quantum}) to second order in $p$ and re-summing the resulting series using Eq.~(\ref{eq:sum-rules}) gives $K_0(I,p)\simeq-(\gamma_\ell-\gamma_\tg)Ip+(\gamma_\ell+\gamma_\tg)p^2\Gamma(I)/2$.  This expansion is justified when the temperature is large, $T\gg \omega_\tp$, and thus the bath occupation number $n_\tB\approx T/\omega_\tp$ is also large, $n_\tB\gg 1$.  Using definitions of the rates $\gamma_\ell-\gamma_\tg=2\kappa$ and $\gamma_\ell+\gamma_\tg=2\kappa\coth(\omega_\tp/2T)\simeq 4\kappa T/\omega_\tp$ in the high temperature limit, one retrieves the classical effective Hamiltonian~(\ref{K0classical}).

As in the classical case, in the term $K_1$ that describes the second drive all Fourier modes must be kept.  Using Eq.~(\ref{ActionAngleKeldysh}) dropping $I^\q$ and with $\theta\to\underline\theta$ we obtain
\begin{equation}\label{K1quantum}
K_1(I,p;t)=2\alpha\sum_n\sinh\!\left(\frac{np}{2}\right)c_n(I)\, e^{-\im(n\underline\theta(I)-\nu t-\varphi_\td)}+\mathrm{c.c.}
\end{equation}
The classical result (\ref{K1classical}) can again be found by expansion in the auxiliary variable $p$. Indeed, the first order expansion of the sinh function leads to Eq.~(\ref{K1classical}). 

Equations (\ref{IpTheory}), (\ref{K0quantum}) - (\ref{K1quantum}) provide the full  description of a weakly dissipative quantum  parametric oscillator, including the effects of the spectroscopic drive, $\omega_\mathrm{d}$. Whereas expanding the effective Hamiltonian, $K=K_0+K_1$ to the second order in $p$ fully recovers the classical results, the opposite transition, from classical to quantum, is far less obvious. This is due to exponential, $\e^{-np}$  and  $\sinh(np/2)$, functions in the quantum version. Upon quantization they represent shift operators of the action $I$ by an integer $n$, reflecting the underlying Bohr-Sommerfeld quantization of the action.

\section{Phase Slip Rate}\label{III}

We now turn to the computation of the phase slip rate.  Without the  spectroscopic drive the PDO respects parity symmetry $\phi\to-\phi$.
The two classical ground states $\phi=\pm\phi_0$ spontaneously break this symmetry.
At long times the symmetry is restored by rare phase slips events, in which the system traverses from the vicinity of one of the ground states in the phase space to the vicinity of another over the quasi-energy barrier.
This results in a bimodal probability distribution over both wells of Fig.~\ref{fig:H0Surface} at long times.

We will consider the case where the frequency detuning $|\nu|$ exceeds the phase-slip rate $W_\mathrm{ps}$ of the switching events. In this case the rates of switching between the two states are  equal, so we do not distinguish from which state the switching occurs.
Within the path integral formalism, the phase-slip rate $W_\mathrm{ps}$ can be computed using the saddle point approximation,
\begin{equation}\label{tswitch}
W_\mathrm{ps}\simeq C\,\e^{\im S_\inst}
\end{equation}
where the prefactor $C$ has only a sub-exponential dependence on the parameters and $S_\inst$ is the action evaluated for an instanton trajectory that connects the classically stable state and the saddle point $\phi=0$ of the classical fluctuation-free trajectory. For switching from the state $\phi_0$, the boundary conditions are $\phi_\mathrm{inst}(-\infty) \to \phi_0$ and $\phi_\mathrm{inst}(\infty)\to 0$, whereas for switching from the state $-\phi_0$ we have $\phi_\mathrm{inst}(-\infty) \to -\phi_0$.
In the weak damping limit, in the both cases the instanton is a trajectory that starts at the bottom of the well at $I=0$ and activates to the top of the quasi-energy barrier, $I_\mathrm{top}$.
The computation of the switching rate exponent thus reduces to identifying such an instanton.

In the language of open quantum systems, these phase-slips lead to the dynamical restoration of {\em weak} parity symmetry in the sense of Refs.~\cite{VAlbertSym&Conserv,VAlbertSymBreakErrorCorrection}.
The problem of symmetry restoration of {\em strong} parity symmetry was considered in \cite{SrongSymInstantons}, which studied a $T=0$ quantum PDO with nonlinear friction.
In both cases, the exponent of the instanton action and thus the phase slip rate is related to (in the present context -- real-valued) dissipative gap in the spectrum of the Lindbladian super-operator.
In the classical stochastic setting, one may similarly argue that the phase slip rate characterizes the dissipative gap of the corresponding Fokker-Planck operator.

\subsection{Classical Activation}

The problem of identifying the instanton trajectory requires understanding the classical mechanics described by the effective Hamiltonian Eq.~(\ref{K0classical}).  Being of the standard form for Langevin processes with multiplicative noise, the activation path is a time-reversed version of the fluctuationless relaxation path.  This owes to the presence of an emergent detailed balance in the classical theory of Markovian systems with one dynamical variable, despite being driven out of equilibrium.

\begin{figure}
    \centering
    \includegraphics{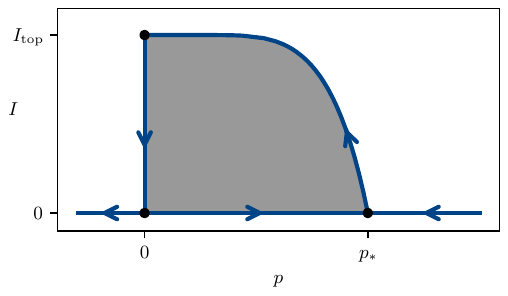}
    \caption{Phase portrait of the effective Hamiltonian $K_0$ from Eq.~(\ref{K0classical}) plotted with $\Delta/2\lambda=0.3$ 
    Bold blue lines show the separatrix trajectories, which are the $K_0=0$ lines. They connect the fixed points, which are shown by dots. The instanton action Eq.~(\ref{SwitchingRate0Classical}) is indicated by the shaded region.}
    \label{fig:classicalInstanton}
\end{figure}

The activation and relaxation trajectories are lines in the semiclassical phase portrait on which $K_0=0$.  The phase portrait is shown in Fig.~\ref{fig:classicalInstanton}.
The corresponding Hamilton equations are:
\begin{subequations}\label{EoMclasscical}
\begin{equation}
\label{eq:I-motion}
\partial_t I=-2\kappa I+\frac{4\kappa T}{\omega_\tp}\,\Gamma(I)\,p,
\end{equation}
\begin{equation}
\partial_tp=2\kappa p-\frac{2\kappa T}{\omega_\tp}\,\partial_I\Gamma(I)\,p^2.
\end{equation}
\end{subequations}
From these, one may identify three invariant lines with $K_0(I,p)=0$, which connect three fixed points.
The $p=0$ line corresponds to the fluctuationless relaxation, along which the evolution of the action is $\partial_tI_\mathrm{rel}=-2\kappa I_\mathrm{rel}$.  This trajectory connects the  saddle point ($p=0$ and $I=I_\mathrm{top}$) to the bottom of the well ($p=0$ and $I=0$).
The $I=0$ line also defines an invariant subspace, along which positive $p$ flows from $p=0$ to a second fixed point at $p_*=\omega_\tp/(T\partial_I\Gamma(0))$.
The third line is the instanton path, which connects the fixed point at $p=p_*$ and $I=0$ to the top of the well $p=0$ and $I=I_\mathrm{top}$, along which both coordinates are non-zero.  Motion along this line describes the activation process.  It is defined implicitly through
\begin{equation}\label{pInstClassical}
p_\inst(I_\inst) =\frac{\omega_\tp}{T}\frac{I_\inst}{\Gamma(I_\inst)}.
\end{equation}
Along such instanton path the action evolves according to $\partial_tI_\inst=2\kappa I_\inst$, cf. Eq.~(\ref{eq:I-motion}), realizing the time-reversed version of the relaxation trajectory. In fact, for an underdamped thermal equilibrium  system this trajectory was found by Kramers \cite{Kramers1940} by solving the Fokker-Planck equation.

Thus, following Eq.~(\ref{tswitch}), the phase-slip rate exponent is given by
\begin{equation}\label{SwitchingRate0Classical}
\im S_\inst=-\int\!\dif t\,p_\inst \partial_tI_\inst=-\frac{\omega_\tp}{T}\int\limits_0^{I_\mathrm{top}}\! \dif I\, \frac{I}{\Gamma(I)}.
\end{equation}
A closed-form analytical result is hardly possible, but it is straightforward to numerically evaluate this expression.
The parametric dependence of the instanton action can be further simplified by noting that the problem can be made dimensionless through a rescaling of the quasi-energy $E\to(2\lambda^2/g)\tilde E$ and time $t\to(1/2\lambda)\tilde t$.  With this, the action and its auxiliary partner are rescaled as $I\to(\lambda/g)\tilde I$ and $p(I)\to\tilde p(\tilde I)$.  The final result is
\begin{equation}
\label{ClaasicalRate}
\im S_\inst=-\frac{\omega_\tp\lambda}{T g}\, R(\Delta/2\lambda)
\end{equation}
where $R=\int\dif\tilde I\, \tilde p_\inst(\tilde I)$ is a dimensionless constant of order unity whose value depends only on the detuning parameter, $\Delta$.
This together with Eq.~(\ref{SwitchingRate0Classical}) concur with results reported in Ref.~\cite{ClassicalActivation}.

\subsection{Quantum activation}

As discussed above, the quantum problem matches the classical problem in the limit of large $T$.  They disagree, however, in the opposite limit.  Equation~(\ref{ClaasicalRate}) predicts that the switching rate goes to zero as $T\to0$.  In contrast, the quantum result shows that it saturates to a finite value.  The finite rate of phase slips at zero temperature is driven by the quantum fluctuations of the bath and is known as quantum activation \cite{QuantumActivation}.

The quantum instanton is given by the $K_0(I,p)=0$ lines of the Keldysh Hamiltonian, Eq.~(\ref{K0quantum}).
The phase portrait is similar to the classical theory in that it consists of three of these lines connecting the fixed points, which generally have the same meaning as in the classical theory.
The $p=0$ line flowing from $I=I_\mathrm{top}$ to $I=0$ describes fluctuationless relaxation, along which the time evolution may be obtained by re-summing the Fourier series of $\partial_pK_0|_{p=0}$.  The result is the same as in the classical theory, $\partial_tI_\mathrm{rel}=-(\gamma_\ell-\gamma_\tg)I_\mathrm{rel}=-2\kappa I_\mathrm{rel}$.
The $I=0$ line connects the bottom of the well to a finite $p$ fixed point $p_*$ with a shifted valued compared to the classical theory, an exact expression for which can be found in Appendix \ref{appB}.

The identification of the instanton line is made easier using an alternative representation for $K_0=\gamma_\ell K_\ell+\gamma_\tg K_\tg$ obtained by re-summing the Fourier series,
\begin{equation}\label{K0quantum_Integral}
K_{\ell,\tg}(I)\!=\!\!\int\!\frac{\dif\theta}{2\pi}\big(\underline{\bar\phi}(\theta\pm\im p/2;I)\underline\phi(\theta\mp\im p/2;I)-\underline{\bar\phi}(\theta;I)\underline\phi(\theta;I)\big)
\end{equation}
In this representation one may immediately identify a $K_0=0$ line with $-\im p(I)=\omega(I)\mathrm{Im}(t_2(I))$ where $t_2(I)$ is the complex period of the elliptic function solutions $\underline\phi(t;I)$.  Such a choice annihilates the integrands of both $K_{\ell,\tg}$.  This determines the tunneling rate of the closed system with the Hamiltonian Eq.~(\ref{HRWA}).
It turns out that there is always another $K_0=0$ line along which the action accrues a smaller value.  This trajectory, which corresponds to quantum activation, leads to an exponentially larger switching rate compared with the coherent tunneling process and is thus the primary mechanism responsible for phase slips.

\begin{figure}
    \centering
    \includegraphics{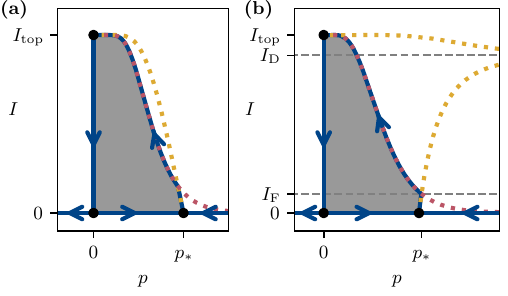}
    \caption{Phase portraits of the quantum Hamiltonian $K_0(I,p)$, Eq.~(\ref{K0quantum}) for $\Delta/2\lambda=0.3$ (a) and $\Delta/2\lambda=-0.6$ (b). Bold blue lines show the separatrix trajectories in the $T\to 0$ limit. They are the $K_0(I,p)=0$ lines, which connect the fixed points, shown by bold dots. The instanton action Eq.~(\ref{SwitchingRate0Quantum}) is indicated by the shaded region. The red dashed lines show the path $p_<(I)$ and the yellow dashed lines show the $T=0$ instanton. Note that this instanton trajectory diverges in panel (b) at $I=I_\tD(\Delta/2\lambda)$.  This is a generic feature of the $T=0$ instanton when $\Delta/2\lambda\leq0$. The singularity of $p_\inst(I)$ is integrable, resulting in a finite switching rate. This singularity is no longer present for $T\to0$ because $I_\tD>I_\tF$ for all values of $\Delta/2\lambda$.}
    \label{fig:quantumInstanton}
\end{figure}

We first consider the instanton line for zero temperature.  This corresponds to $\gamma_\tg=0$ and thus $K_0=K_\ell$.
The instanton line must annihilate the integrand of Eq.~(\ref{K0quantum_Integral}).  This turns out to be possible for $p_\inst<\omega\mathrm{Im}(t_2)$ due to a certain quasi-periodicity of the classical solutions $\underline\phi$, which obey the condition $\underline\phi(t-t_\tQ)=\underline{\bar\phi}(t)$.  The quasi-period time $t_\tQ(I)$ is purely imaginary and always smaller in magnitude than $\mathrm{Im}(t_2(I))/2$; the functional form of $t_\tQ$ and further details are discussed in Appendix \ref{appA}.  This gives the instanton implicitly as a function of action
\begin{equation}\label{pinstT=0}
p_\inst^{(T=0)}(I_\inst)=-2\im\omega(I_\inst)t_\tQ(I_\inst).
\end{equation}
Feeding this back into the equation of motion for $I$ gives $\partial_tI_\inst=\gamma_\ell I_\inst=2\kappa I_\inst$, which is exactly the time-reversed relaxation trajectory.  This may be traced back to the fact that at $T=0$ detailed balance emerges in the quantum system, similar to the classical result \cite{QuantumActivation}.
The $T=0$ phase portrait of $K_0$ along with the corresponding instanton trajectories are shown in Fig.~\ref{fig:quantumInstanton}.
The switching rate exponent is given by substituting this result into the action,
\begin{equation}\label{SwitchingRate0Quantum}
\im S_\inst=-2\int\limits_0^{I_\mathrm{top}}\dif I\,\omega(I)|t_\tQ(I)|\equiv-\frac{\lambda}{g}\, R^{(T=0)}(\Delta/2\lambda),
\end{equation}
where $R^{(T=0)}$ is a dimensionless constant that depends only on the dimensionless ratio $\Delta/2\lambda$.

The finite temperature problem differs in a qualitative way from the $T=0$ problem in that the instanton $p_\inst$ comes from a cancellation between $K_\ell$ and $K_\tg$ rather than annihilating their integrands individually.
The exact form of the finite temperature instanton is not easy to deduce analytically, however it may be obtained in certain limits.  The resulting phase slip time interpolates between the $T\to\infty$ limit, which matches the classical result, and the $T\to0$ limit.
This limit is not always equivalent to the $T=0$ result.
Such fragility \cite{QuantumActivation} of the zero temperature result indicates the breakdown of the semiclassical approximation for some very small temperature scale at which a crossover between the $T=0$ and $T\to0$ results occurs.

The origin of the zero temperature fragility can be traced to the series representation of the Keldysh Hamiltonian, Eq.~(\ref{K0quantum}).  Following Eq.~(\ref{Wn}) for $T=0$, $\gamma_\tg=0$ and so the transition rates are $W_n\propto|c_n|^2$, which fall off as $\exp(-np_<)$ for $n\gg 1$ and $\exp(-|n|p_>)$ for $-n\gg 1$ (expressions for $p_\gtrless \equiv p_\gtrless(I; \Delta/2\lambda)$ are given in Appendix~\ref{appA}). Therefore the series $\sum_n W_n\exp(-np)$ has an asymmetric radius of convergence $-p_< <p <p_>$.  It is important that $p_<(I;\Delta/2\lambda)\neq p_>(I;\Delta/2\lambda)$ in a broad range of $I$ and $\Delta/2\lambda$.
For any finite $T>0$ the additional factor of $|c_{-n}|^2$ in $W_n$ changes the radius of convergence to be symmetric, $-\tmin(p_>,p_<)<p<\tmin(p_>,p_<)$.

It turns out that in a certain region of $\Delta/2\lambda$ there exists a range $I>I_\mathrm{F}(\Delta/2\lambda)$ where the zero temperature instanton from Eq.~(\ref{pinstT=0}) falls outside the reduced radius of convergence,  $p_<(I;\Delta/2\lambda)<p^{(T=0)}_\inst(I;\Delta/2\lambda)$.
In this range the limit $T\to 0$ brings the instanton to the edge of the radius of convergence, so that $p_\inst^{(T\to0)}=p_<$, or in total
\begin{equation}\label{pinstTto0}
p_\inst^{(T\to0)}=\begin{cases}p_\inst^{(T=0)},&I\in[0,I_\tF)\\p_<,&I\in(I_\tF,I_\mathrm{top})\end{cases}
\end{equation}
where $I_\tF$ defines the value of the action above which the zero temperature result is fragile.
This piecewise form of the $T\to0$ limit of the instanton is shown in Fig.~\ref{fig:quantumInstanton}.
The fragility of the zero temperature result indicates the breakdown of the semiclassical approximation for some very small temperature scale, at which a crossover between the $T=0$ and $T\to0$ results occurs \cite{QuantumActivation}.  For a resonantly driven oscillator the crossover temperature was investigated in Ref.~\cite{QuantumCriticalTemperature}.

\section{Logarithmic Susceptibility}\label{IV}

We now turn to the computation of the logarithmic susceptibility (LS).  In the presence of a finite $\alpha$ -- the spectroscopic signal amplitude -- there is an exponential correction to the phase slip rate.  The correction to the exponent may be computed perturbatively $S_\inst\simeq S_\inst^{(0)}+S_\inst^{(1)}$ to first order in $\alpha$.

This problem was recently solved for the case when the linear drive is at half resonance with the parametric drive $\omega_\td=\omega_\tp$ \cite{QuantumSymmetryLifting,QuantumSymmetryLiftingZeroT}.  In this case $\nu=0$ and so the rotating-wave approximation results in a time-independent theory of switching in a tilted double well in the phase space.
The technical approach in \cite{QuantumSymmetryLifting} was to work in the unperturbed action-angle variables and absorb the tilting of the double well into perturbative corrections to the Fourier coefficients in Eq.~(\ref{ActionAngle}).  The perturbative correction to the instanton $p_\inst=p_\inst^{(0)}+\alpha p_\inst^{(1)}$ is then obtained by solving $K_0+K_1=0$ for $p_\inst^{(1)}(I)$ at order $\alpha$.
It results in an asymmetric correction to the phase slip rate, so that switching out of the shallower well into the deeper well becomes exponentially more likely.
The results discussed there can be extended to the regime of very small $\nu$ by replacing $\varphi_\td\to\varphi_\td+\nu t$.  This is valid in the adiabatic limit $\nu \ll W_\mathrm{ps}$, in which there are many phase-slip events over a single period of modulation of the wells.

Here, as indicated earlier, we consider the regime where $\nu = \omega_\td -\omega_\tp$ is large enough so that the modulation of the double well is fast compared to the phase slip rate, such that many modulation periods occur in-between successive  phase slips. This is qualitatively different from the adiabatic limit. Over timescales large compared to $1/\nu$ the symmetry of the problem is restored and switching rates between the wells are again symmetric. However, the spectroscopic drive still leads to an exponentially strong increase in the switching rates. Whenever the frequency $\nu$ or its integer multiples resonate with the motion at the bottom of the wells this increase becomes particularly large.

On a technical level, this problem is unavoidably time-dependent and so requires an alternative approach from that used for the half resonant linear drive.
Such an approach is provided by the standard machinery for computing time-dependent perturbations to the instanton action (see Chapter 4 of \cite{KamenevBook} for a review).  The first-order correction is given by evaluating the perturbing action $\im S_1=\int \dif t\,K_1(I,p;t)$ along the unperturbed instanton trajectory,
\begin{equation}\label{S1}
\im S_\inst^{(1)}=-\min_{t_0}\int\dif t\,K_1\big(I_\inst(t-t_0),p_\inst(t-t_0);t\big),
\end{equation}
where the minimization is taken over the central time $t_0$ of the bare instanton trajectory.  The time-dependence of the perturbation breaks the time translation invariance of the unperturbed result, pinning the instanton center to the time where the switching event is the most likely to occur \cite{LSoriginal}.
This formula is applicable to both the classical and quantum problems, which are treated on the same footing in the action formulation.  For both, this can be understood as the Melnikov theorem \cite{KamenevBook} applied over the effective phase space guaranteeing that the perturbed instanton remains on the stable manifold of the saddle point of the fluctuation-free classical motion.

Using the specific form of $K_1$, either the classical result from Eq.~(\ref{K1classical}) or the quantum result from Eq.~(\ref{K1quantum})
, Eq.~(\ref{S1}) becomes
\begin{equation}\label{ActionClassicalLS}
\im S_\inst^{(1)}=-\alpha\,\underset{t_0}{\mathrm{min}}\left\{ \int\!\dif t\ \chi(t-t_0)\, e^{\im(\nu t+\varphi_\td)}+\mathrm{c.c.}\right\}, 
\end{equation}
where, following Ref.~\cite{LSoriginal},  $\chi(t)$ defines the logarithmic susceptibility. The LS is essentially a response function for the switching rate exponent. Its computation for the classical and quantum problems are presented below.

\subsection{Classical Logarithmic Susceptibility}

The computation of the LS for a classical PDO was originally performed in Ref.~\cite{ClassicalLS}.
The method employed there in the analysis of the LS in the weak damping limit is to express the LS as a sum over different harmonic contributions and argue that, for certain values of $\nu$, the resonant harmonic gives the dominant contribution, which can be found by the steepest descent. The strongest response comes from the first harmonic with $n=1$, when $\nu \lesssim \omega_\tmin$.  When $\nu$ exceeds $\omega_\tmin$, the external drive can no longer resonate with the dominant harmonic of the intrawell motion, causing a sharp drop off in the LS. Resonance with sub-leading harmonics with $n>1$ is still possible however, but the response is weaker.  Increasing $\nu$ produces a series of drop-offs in the LS corresponding to $\nu$ exceeding integer multiples of $\omega_\tmin$.
For the classical theory, this leads to resonant peaks in the LS for $\nu$ near integer multiples of $\omega_\tmin$.

Here, we compute the LS by using the form of $K_1$ from Eq.~(\ref{K1classical}), which appears in the form of a series.  This can be decomposed into harmonics $\chi(t)=\sum_n\chi_n(t)$. Additionally, note that Eq.~(\ref{ActionClassicalLS}) is just the Fourier transform of $\chi$, minimized over an overall phase. After shifting the time argument and absorbing the phase shift $\varphi_\td$ into the minimization, the correction to the switching rate exponent becomes:
\begin{equation}\label{ActionChiFT}
\im S_\inst^{(1)}=-\alpha\, \underset{t_0}{\mathrm{min}}\bigg\{ \e^{\im\nu t_0}\sum_n\chi_n(\nu)+\mathrm{c.c.}\bigg\},
\end{equation}
with $\chi_n(\nu)$ the Fourier transforms of $\chi_n(t)$,
\begin{equation}\label{ChiFT}
\chi_n(\nu)=\!\int\!\dif t\ \e^{-\im(n\underline\theta(t)-\nu t)}\, c_n\big(I_\inst(t)\big) n p_\inst(t) , 
\end{equation}
where $\underline\theta(t)=\int^t \dif t \, \omega(I_\inst(t))$, while $\partial_t \underline\theta(t) = \omega(I_\inst(t))$ 
and $\partial_t^2 \underline\theta(t) = \partial_I \omega(I_\inst)\partial_t I_\inst = 2\kappa I_\inst \partial_I \omega(I_\inst)$, where the instanton equation of motion, $\partial_tI_\inst=2\kappa I_\inst$ was used.  
The integral in Eq.~(\ref{ChiFT}) can be evaluated via the saddle point approximation, where all terms outside the exponent are treated as slow. This brings the saddle point condition $n\omega(I_\inst(t))=\nu$, 
which selects a certain resonant time $t_\mathrm{r} < \log(I_\mathrm{top}/2\kappa)$, along the instanton path $I_\inst(t)=\e^{2\kappa t}$ and the corresponding action $I=I_\inst(t_\mathrm{r})$, such that $0<I<I_\mathrm{top}$, where the saddle point condition is satisfied. The corresponding LS, up to an overall phase factor, is given by  
\begin{equation}\label{LSharmonicClassical}
\chi_n(\nu)\simeq\frac{\omega_\tp}{T}\,
\sqrt\frac{\pi\,|n|\, I}{\kappa\, |\partial_I\omega(I)|}\, 
\frac{c_n(I)}{\Gamma(I)} \ \Bigg\vert_{n\omega(I)=\nu},
\end{equation}
where Eq.~(\ref{pInstClassical}) was employed.

Function $\omega(I)$ is  positive and monotonically decreasing from $\omega(0) =\omega_\mathrm{min}$ down to $\omega(I_\mathrm{top})=0$. Therefore for a given  value of $\nu$ the saddle point condition is only satisfiable for sufficiently large values of $n$.  For $(|m|-1)\omega_{\tmin}<|\nu|\leq|m|\omega_{\tmin}$, one must choose $|n|\geq|m|$. 
When the saddle point condition is not met, the functions $\chi_n(\nu)$ cannot be evaluated by a saddle point approximation and provide only small background contributions to $\chi_n(\nu)$.

For a given value of $\nu$, it turns out that the harmonic $\chi_n(\nu)$ corresponding to the smallest value  $n=m$, for which the saddle point condition can be satisfied, gives the dominant contribution to the LS. This is a consequence of the exponential decrease of the Fourier coefficients, $c_n$, with $n$, see Appendix \ref{appB}.


Because of this, a good approximation is provided by keeping only the harmonic corresponding to the smallest value of $n$ satisfying the saddle point condition.
Thus, for $(n-1)\omega_\tmin< \nu< n\omega_{\tmin}$  Eq.~(\ref{ActionChiFT}) results in the correction to the switching rate exponent,
\begin{equation}\label{S1chi}
\im S_\inst^{(1)} \simeq 2\alpha |\chi_n(\nu)|,
\end{equation}
where the minimization over $t_0$ has eliminated the overall phase.
This expression is peaked near the upper boundary of the frequency interval $\nu \simeq n\omega_\tmin$, where $I\simeq 0$. 

\begin{figure}
    \centering
    \includegraphics{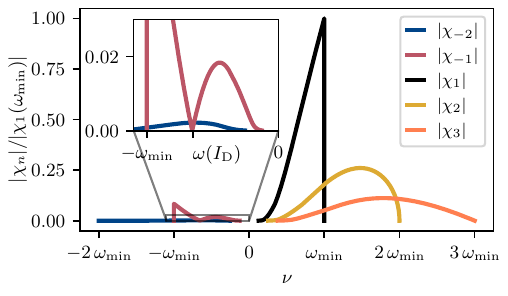}
    \caption{Classical phase slip rate logarithmic susceptibility as a function of reduced spectroscopic frequency, $\nu =\omega_\td-\omega_\tp$ according to Eq.(\ref{LSharmonicClassical}) for $\Delta/2\lambda = -0.4$. For $\Delta<0$ and $n<0$, at $n\omega(I_\tD) = \nu$, see Fig.~\ref{fig:quantumInstanton}b,  the corresponding $\chi_n$ becomes zero, here highlighted for $\chi_{-1}$ in the inset. }
    \label{fig:Classical_Log_Susc}
\end{figure}

The shape of each resonant peak can be extracted from Eq.~(\ref{LSharmonicClassical}) by putting $|\nu|=|n|\omega_{\tmin}-\delta\nu$ and expanding to leading order in $\delta\nu$.  The saddle point condition for $\chi_n$ then becomes $\omega(I)=\omega_{\tmin}-\delta\nu/|n|$.
This corresponds to early times along the instanton path and therefore small action 
$I\simeq \delta\nu/|n\omega_I|$, where
\begin{equation}
\omega_I\equiv \partial_I\omega(I)\vert_{I=0}=-g\, \frac{\lambda-\Delta/8}{\lambda-\Delta/2}.
\end{equation}
This, along with the limiting forms of $\Gamma$ and $c_n$ for small $I$ are discussed in Appendix \ref{appB}.
In particular, $\Gamma(I\to 0) \propto I$ and $c_1\propto\sqrt{I}$, which gives for the dominant resonance, $n=1$,
\begin{equation}\label{LSharmonicResonanceClassical}
|\chi_1(\nu)|\underset{\delta\nu\ll |\omega_I|}{\simeq} a_1 \frac{\omega_\tp}{T}\sqrt\frac{\pi}{\kappa|\omega_I|}\,\left(1+b_1\frac{\delta\nu}{|\omega_I|}\right)\Theta(\omega_{\tmin}-\nu),
\end{equation}
where $a_1$ and $b_1$ are dimensionless constants of order unity.
For sub-leading resonances with $n\neq 1$, one obtains
\begin{equation}\label{LSharmonicResonanceClassical-n}
|\chi_n(\nu)| \underset{\delta\nu\ll |\omega_I|}{\simeq} a_n \frac{\omega_\tp}{T}\sqrt\frac{\pi}{\kappa|\omega_I|}\,\Bigg|\frac{\delta\nu}{n\omega_I}\Bigg|^{\frac{|n|-1}{2}}\Theta(|n|\omega_{\tmin}-|\nu|).
\end{equation}
The Heaviside theta function $\Theta$ in both expressions accounts for the inability to satisfy the saddle point condition of $|\nu|>|n|\omega_\tmin$ (for a fixed $n$). As a result, each $n$-peak features the sharp, non-analytic drop-off of the LS for $|\nu|>|n|\omega_\tmin$.
This reproduces the main results of Ref.~\cite{ClassicalLS} on the LS in the weak-damping limit.

The precise form of the constants $a_n$ may be obtained from the $I\to0$ limiting forms of $\Gamma$ and $c_n$ discussed in Appendix \ref{appB}.  For example, the height of the dominant peak of the LS is set by $a_1$, which has the explicit form:
\begin{equation}\label{a1}
a_1=\sqrt\frac{\omega_\tmin(\bar\omega+\omega_\tmin)}{2\bar\omega^2},
\end{equation}
with $\bar\omega=4\lambda-\Delta$.

Interestingly, for $\Delta<0$ there is a narrow region of frequencies for which the leading-order harmonic does not provide the dominant contribution to the LS.  This occurs close to the frequency $\nu_n = n \omega(I_\mathrm{D})$, at which $c_n$, and consequently also $\chi_n$, vanish identically when $n<0$.
The value of $I_\mathrm{D}$ corresponds to the point where the zero temperature quantum instanton in the absence of the spectroscopic drive diverges, see also Fig.~\ref{fig:quantumInstanton}b (a specific expression is given in Appendix \ref{appA}).
In the vicinity of $\nu_n$ the next higher harmonics $\chi_{n-1}$ give the dominant contribution to $\im S_\inst^{(1)}$. For $n=-1$ this behavior is shown in the inset of Fig.~\ref{fig:Classical_Log_Susc}.


A significant feature of the LS is the scaling with the dissipation strength $\chi\sim\kappa^{-1/2}$.  Because this result is valid in the limit of weak dissipation, this scaling provides a significant enhancement to the magnitude of the LS.  As will be shown below, this scaling persists in the quantum theory at $T=0$.  This should be contrasted with the results known for both the classical \cite{ClassicalLS} and quantum \cite{QuantumSymmetryLifting,QuantumSymmetryLiftingZeroT} PDO with a commensurate second drive $\omega_\td=\omega_\tp$, breaking the symmetry between the two wells, for which the LS does not scale with the dissipation strength, $\chi\propto\kappa^0$.
This implies that the non-commensurate  second drive $\omega_\td\neq\omega_\tp$ produces a parametrically larger response, resulting in a bigger enhancement of the phase-slip rate.
It can be understood to be a result of the resonance of the second drive with the intrawell motion.

\subsection{Quantum Logarithmic Susceptibility}

Equation~(\ref{LSharmonicResonanceClassical}) predicts the LS to diverge in the limit of $T\to0$, which would signal the breakdown of the perturbative expansion for sufficiently small $T$.  Like the unperturbed switching rate, quantum effects curtail this divergence of the LS and give a finite result down to zero temperature.   

The same methodology described above for the classical result in the weak-damping limit can be employed for the quantum problem.  The leading correction to the switching rate exponent is provided by Eq.~(\ref{S1}), where the Fourier transform of the harmonics $\chi_n$ are obtained from $K_1$ given by the quantum theory, Eq.~(\ref{K1quantum}),
\begin{equation}
\chi_n(\nu)=\!\int\!\dif t\e^{-\im(n\underline\theta(t)-\nu t)}c_n\big(I_\inst(t)\big)2\sinh\left(\frac{np_\inst(t)}{2}\right)
\end{equation}
The integral may be evaluated via saddle point approximation, with all terms outside the exponent treated as slow,
\begin{equation}\label{LSharmonicQuantum}
\chi_n(\nu)=
\frac{\sqrt{2\pi}\, c_n(I)\,  2\sinh(np_\inst(I)/2)}{\sqrt{n\partial_I\omega(I)\sum_{n'}n'W_{n'}(I)\e^{-n'p_\inst(I)}}}\Bigg\vert_{n\omega(I)=\nu},
\end{equation}
where $p_\inst(I)$ is given for $T\to0$ by Eqs.~(\ref{pinstT=0}), (\ref{pinstTto0}), see also Fig.~\ref{fig:quantumInstanton}. 
This gives the quantum analogue to Eq.~(\ref{LSharmonicClassical}).

Just as for the classical theory, for a given value of $\nu$ the harmonic $\chi_n$ corresponding to the smallest value of $n$ for which the saddle point condition is satisfiable provides the dominant contribution to the switching rate.  Despite the different form of Eq.~(\ref{LSharmonicQuantum}) compared to Eq.~(\ref{LSharmonicClassical}), the harmonics for $n>\nu/\omega_\tmin +1$ are exponentially suppressed in $n$, allowing to keep the only one $n$, such that  $\nu/\omega_\tmin < n<\nu/\omega_\tmin+1$.   
The leading harmonic $\chi_n(\nu)$ displays a sharp, non-analytic drop for $\nu$ near integer multiples of $\omega_\tmin$.

This may be obtained by putting $|\nu|=|n|\omega_\tmin-\delta\nu$ and expanding to leading order in $\delta\nu$.  Like in the classical problem this requires evaluating Eq.~(\ref{LSharmonicQuantum}) for $I\to0$.  In this limit the instanton is replaced by its value at the fixed point $p_\inst(I\to0)=p_*$. 
The sum in the denominator originates from $\partial_tI_\inst(I)$ from the saddle point condition; for early times along the instanton path $\partial_tI_\inst(I)\simeq2\kappa I$ even away from zero temperature (see Appendix \ref{appC}) so this term cancels as in the classical problem.
Putting everything together gives the quantum analogue of Eqs.~(\ref{LSharmonicResonanceClassical}) and (\ref{LSharmonicResonanceClassical-n}) for the dominant harmonic and sub-leading harmonics,
\begin{subequations}\label{LSharmonicResonanceQuantum-all}
\begin{equation}\label{LSharmonicResonanceQuantum}
|\chi_1(\nu)|\underset{\delta\nu\ll|\omega_I|}{\simeq} A_1\sqrt\frac{\pi}{\kappa|\omega_I|}\, \left(1+B_1\frac{\delta\nu}{|\omega_I|}\right)\Theta(\omega_{\tmin}-\nu),
\end{equation}
\begin{equation}\label{LSharmonicResonanceQuantum-n}
|\chi_n(\nu)|\underset{\delta\nu\ll|\omega_I|}{\simeq}A_n\sqrt\frac{\pi}{\kappa|\omega_I|}\, \Bigg|\frac{\delta\nu}{n\omega_I}\Bigg|^{\frac{|n|-1}{2}}\Theta(|n|\omega_{\tmin}-|\nu|).
\end{equation}
\end{subequations}
The dimensionless constants $A_n$ have a complex dependence on both the bifurcation parameter $\Delta$ and the temperature through $\gamma_{\tl,\tg}$ (see Appendix \ref{appC}).
Crucially, they remain finite for $T\to0$, for example,
\begin{equation}\label{A1}
A_1=\sqrt\frac{2\omega_\tmin(\bar\omega+\omega_\tmin)}{(2n_\tB+1)^2\bar\omega^2-\omega_\tmin^2},
\end{equation}
with $\bar\omega=4\lambda-\Delta$.
The $T\to0$ limit is given by $n_\tB\to0$, which remains finite. Yet, $A_1$ may reach a rather large value in this limit. Indeed, if $\lambda\gg\Delta$, one finds from Eq.~(\ref{A1}) $A_1 =8\lambda/\Delta \gg 1$. 
At high temperature $A_1 \simeq a_1\omega_\tp/T$, and $B_1\simeq b_1$, putting Eq.~(\ref{LSharmonicResonanceQuantum}) in agreement with Eq.~(\ref{LSharmonicResonanceClassical}). 
As a result, the quantum limit is reached at the temperature 
\begin{equation}
    T\lesssim T^* = \frac{\omega_\tp}{\log(\omega_\tmin/\Delta) },
\end{equation}
suppressed by the entropic logarithm, due to the large density of states. 

Thus the quantum result predicts the sharp drops in the LS for $\nu$ near integer multiples of $\omega_\tmin$ persist down to $T=0$.  In addition, the scaling $\chi\sim\kappa^{-1/2}$ is maintained in the $T=0$ limit.
This can be seen in Fig.~\ref{fig:QLS}, which shows LS harmonics $\chi_n$ for different values of the the detuning $\Delta$ and temperature $T$.
There, the sharp, non-analytic drop of the LS near $n\omega_\tmin$ predicted by Eq.s~(\ref{LSharmonicResonanceQuantum-all}), as well as the finiteness of the dominant harmonic peak Eq.~(\ref{A1}) at $T=0$, are clearly visible.

This result is valid when $\Delta>\lambda$, when there is no fragility and the $T\to0$ limit is continuous.  When $\Delta<\lambda$ the fragility of the $T\to0$ limit signals the breakdown of the single-instanton approximation for very small $T$ and the perturbation theory around such a solution is not reliable.  Indeed, substituting Eq.~(\ref{pinstT=0}) into Eq.~(\ref{LSharmonicQuantum}) when $\Delta<\lambda$ gives harmonic contributions $\chi_n$ which grow in magnitude with increasing $n$, rendering the approximate decomposition of the LS into separate harmonic contributions below each integer multiple of $\omega_\tmin$ invalid.  For $\Delta<0$ the situation is worse, as the divergence of the instanton (see Fig. \ref{fig:quantumInstanton}) manifests as a divergence of $\chi_n$ at $\omega(I_\tD)$.
Thus, eq.~(\ref{LSharmonicQuantum}) should be understood as applying when $\Delta>\lambda$ all the way to $T=0$ or for any $\Delta$ at very small but finite temperatures.
The modification to the rate of phase slips at $T=0$ when $\Delta<\lambda$ must be computed using a different approach, which we leave to future work.

\section{Near Vicinity of the Bifurcation Point}\label{V}

Close to the bifurcation point $\Delta\to2\lambda$, the frequency of the bottom of the well becomes small, $\omega_\tmin\to0$.  Once $\kappa\sim\omega_\tmin$ the dissipation can no longer be regarded as small and the motion becomes overdamped.
When this occurs, there is no longer any coherent motion with which the external drive can resonate.  The LS as a function of $\nu$ has a simple monotonically decreasing form.  The sharp drops near integer multiples of $\omega_\tmin$ discussed in the previous section are absent owing to the lack of resonance.
In this regime, the evolution of the angle coordinate is no longer fast and so the action-angle variables are no longer a useful choice of coordinates and a new approach to calculating the instanton is needed.

A finite value of $\kappa$ shifts the location in phase space of the classically stable attractors.  The two fixed points of Eq.~(\ref{LangevinPDO}) in the absence of noise are $\phi=\pm\phi_0$, where $\phi_0=|\phi_0|\e^{\im\varphi}$ with $|\phi_0|^2=(\sqrt{4\lambda^2-\kappa^2}-\Delta)/g$
and $2\varphi=\pi-\arcsin(\kappa/2\lambda)$.
These fixed points exist so long as $\Delta<\Delta_\tB$, where $\Delta_\tB=\sqrt{4\lambda^2-\kappa^2}$ is the critical detuning, when accounting for a finite dissipation strength.
For small $\kappa$, $\Delta_\tB\simeq2\lambda$ as was used in the preceding sections.
As the dissipationless bifurcation point $\Delta/2\lambda=1$ is approached, the function $H_0$ becomes very flat along the axis parallel to the line in phase space that contains the fixed points $\pm\phi_0$ and $0$.
This line is rotated off the imaginary axis for finite values of $\kappa$.
The real coordinate corresponding to this direction thus evolves slowly, while the coordinate corresponding to the perpendicular direction dissipates at a fast rate.
The switching rate can be well-approximated by first integrating out the fast perpendicular coordinate.

In the classical setting, this is conveniently achieved by going to the path integral description for the original Langevin equation Eq.~(\ref{LangevinPDO}), which is obtained by following a similar procedure to Eq.~(\ref{ZItheta}) for the complex coordinates $\phi$ and $\bar\phi$.  The delta functions that contain the Langevin equations for $\phi$ and $\bar\phi$ can be written as integrals over complex auxiliary fields, which will be respectively denoted $\bar\phi^\q$ and $\phi^\q$.  After integrating out the noise, the partition function is then a path integral $Z=\int\fdif\bar\phi^\alpha\fdif\phi^\alpha\exp(\im S[\bar\phi^\alpha,\phi^\alpha])$ over the complex physical and auxiliary fields $\phi^\alpha=[\phi\ \phi^\q]$, where the action is $S=S_0+S_1$,
\begin{subequations}\label{ActionComplex}
\begin{multline}
S_0[\bar\phi^\alpha,\phi^\alpha]=\int\dif t\Big(\bar\phi^\q(\im\partial_t-\Delta+\im\kappa)\phi
+\bar\phi(\im\partial_t-\Delta-\im\kappa)\phi^\q\\-2\lambda(\bar\phi^\q\bar\phi+\phi^\q\phi)
+\frac{2T\kappa}{\omega_\tp}\,\bar\phi^\q\phi^\q-\frac{g}{2}(\bar\phi^\q\phi+\bar\phi\phi^\q)\bar\phi\phi\Big),
\end{multline}
\begin{equation}
S_1[\bar\phi^\alpha,\phi^\alpha]=-\alpha\int\dif t\big(\bar\phi^\q\e^{-\im(\nu t+\varphi_\td)}+\phi^\q\e^{\im(\nu t+\varphi_\td)}\big).
\end{equation}
\end{subequations}

The decomposition into slow and fast modes is achieved by decomposing $\phi$ into a set of two real variables $\phi=\im\e^{-\im\varphi}(X+\im Y)$ and $\phi^\q=\e^{-\im\varphi}(P_Y-\im P_X)$, where the integration contour of the auxiliary variables is in the imaginary direction $P_X,P_Y\in\im\R$.
The resulting unperturbed fluctuationless $P_X=0=P_Y$ equations of motion for the physical fields are then
\begin{subequations}
\begin{equation}
\partial_tX=\Big(\Delta-\Delta_\tB+\frac{g}{2}(X^2+Y^2)\Big)Y,
\end{equation}
\begin{equation}
\partial_tY=-2\kappa Y-\Big(\Delta+\Delta_\tB+\frac{g}{2}(X^2+Y^2)\Big)X
\end{equation}
\end{subequations}
For $\Delta$ close to $\Delta_\tB$, the $X$ coordinate evolves slowly while the $Y$ coordinate is fast.
Therefore $Y$ and $P_Y$ can be integrated out at the linear level to obtain an effective theory for the slow variables $X$ and $P_X$.
This is achieved by neglecting the non-linear terms and fluctuations of $Y$ and keeping only the term $P_Y(2\kappa Y+(\Delta+\Delta_\tB)X)$.
Integrating out $Y$ and $P_Y$ then imposes the replacement $Y\to-(\Delta+\Delta_\tB)X/2\kappa$.

To make this more precise, introduce $\Delta=\Delta_\tB-2\lambda\epsilon$ where $\epsilon$ is a small dimensionless parameter that measures distance to the bifurcation.  The coordinate $Y$ is fast in the limit $\epsilon\to0$, so the resulting effective theory for $X$ should keep track of each term order by order in $\epsilon$.
This is conveniently achieved by going to dimensionless coordinates $X=\sqrt{4\lambda\epsilon/g}\,Q$, $P_X=\sqrt{4\lambda\epsilon^3/g}\, P$, $t=\tilde t/2\lambda\epsilon$.
The final result of this procedure is a path integral over two coordinates:
\begin{subequations}\label{ZQP}
\begin{equation}
Z=\int\fdif Q\fdif P\ \e^{\im S[Q,P]},
\end{equation}
\begin{equation}
\im S[Q,P]=-\frac{8\lambda}{g}\epsilon^2\int\dif\tilde t\big(P\partial_{\tilde t}Q-K(Q,P)\big),
\end{equation}
\end{subequations}
where the effective Hamiltonian $K=K_0+K_1$ is given to leading order in $\epsilon$ by:
\begin{subequations}\label{KQP}
\begin{equation}
K_0(Q,P)=-P\,U '(Q)+DP^2,
\end{equation}
\begin{equation}
K_1(Q,P;\tilde t)=\frac{\alpha}{4}\sqrt\frac{g}{\lambda^3\epsilon^3}\cos\big((\nu/2\lambda\epsilon)\tilde t+\varphi_\td+\varphi\big).
\end{equation}
\end{subequations}
Here $U(Q)=-(\Delta_\tB/2\kappa)Q^2+(\Delta_\tB\lambda^2/\kappa^3)Q^4$ and $D=\kappa T/\omega_\tp\lambda$.

This is the effective Hamiltonian for a particle undergoing frictional relaxation in a 1D quartic double-well potential. As such, the activation path is given by $P_\inst(Q)=U'(Q)/D$ and the switching rate is the height of the barrier,
\begin{equation}\label{Sinst0QP}
\im S_\inst^{(0)}=-\frac{8\lambda}{g}\epsilon^2\! \int\limits_{Q_\tmin}^0\! \dif Q\ \frac{U'(Q)}{D}=-\frac{\Delta_\tB\omega_\tp\epsilon^2}{2gT}.
\end{equation}
The instanton solution $Q_\inst(t)$ is the standard time-reversed relaxation path $Q_\inst(t)=Q_\mathrm{rel}(-t)$, where $\partial_tQ_\mathrm{rel}=-U'(Q_\mathrm{rel})$.
It is straight forward to evaluate this directly and substitute the result into $S_1$ and subsequently minimize over the instanton center time $t_0$ to obtain the correction to the switching rate exponent \cite{ClassicalLogSusceptibility}
\begin{equation}\label{Sinst1QP}
\im S_\inst^{(1)}(\nu)=\frac{\alpha\omega_\tp}{2T}\sqrt\frac{\epsilon}{\pi\lambda g}\bigg|\boldsymbol\Gamma\bigg(\frac{1-\im\upsilon}{2}\bigg)\boldsymbol\Gamma\bigg(1+\frac{\im\upsilon}{2}\bigg)\bigg|,
\end{equation}
where $\boldsymbol\Gamma$ is the gamma function and $\upsilon=\kappa\nu/2\lambda\Delta_\tB\epsilon$. Higher order terms in $\epsilon$ deform the shape of the potential $U(Q)$ but the double-well shape is preserved, and so the problem is qualitatively unchanged.


The quantum problem near the bifurcation point can be solved using the same set of ideas.
Beginning from the Keldysh action Eq.~(\ref{ActionAngleKeldysh}), the action is brought to a form similar to Eq.~(\ref{ActionComplex}) by switching to the Keldysh-rotated fields $\phi=(\phi^++\phi^-)/2$ and $\phi^\q=\phi^+-\phi^-$.
The resulting action is identical to Eq.~(\ref{ActionComplex}), except for the prefactor in front of the $\bar\phi^\q\phi^\q$ term, which is changed to $2T\kappa/\omega_\tp\to\kappa(2n_\tB+1)$, and the presence of an additional nonlinear term $-(g/8)(\bar\phi^\q\phi+\bar\phi\phi^\q)\bar\phi^\q\phi^\q$.
The decomposition into slow and fast real variables $X$ and $Y$ can also be used in the quantum setting.  After integrating out $Y$ and rescaling $X$ to the dimensionless variable $Q$, one obtains an effective theory of the same form as Eq.~(\ref{ZQP}).  The corresponding effective Hamiltonian $K=K_0+K_1$ matches the form of Eq.~(\ref{KQP}) at leading order in $\epsilon$, but with $D=\kappa(2n_\tB+1)/2\lambda$.

\begin{figure}
    \centering
    \includegraphics{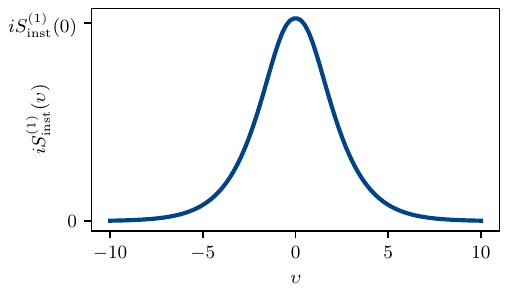}
    \caption{Quantum phase slip rate logarithmic susceptibility as a function of scaled spectroscopic frequency, $\upsilon=\kappa\nu/2\lambda\Delta_\tB\epsilon$.}
    \label{fig:OverdampedS1}
\end{figure}

This leads to similar results to the classical theory, with the main difference being that the quantum result is finite down to $T=0$.  The switching rate is given by Eq.~(\ref{Sinst0QP}) with the  replacement of $2T/\omega_\tp\to  \coth(\omega_\tp/2T)$  
\begin{equation}
\im S_\inst^{(0)}=- \tanh \left(\frac{\omega_\tp}{2T}\right)  \frac{\Delta_\tB\epsilon^2}{g}.
\end{equation}
Similarly, the correction to the switching rate exponent is similar in form to Eq.~(\ref{Sinst1QP})
\begin{equation}
\im S_\inst^{(1)}(\nu)=\alpha\tanh \left(\frac{\omega_\tp}{2T}\right) \sqrt\frac{\epsilon}{\pi\lambda g}\bigg|\boldsymbol\Gamma\bigg(\frac{1-\im\upsilon}{2}\bigg)\boldsymbol\Gamma\bigg(1+\frac{\im\upsilon}{2}\bigg)\bigg|
\end{equation}
In the limit $T\gg\omega_\tp$ one recovers Eq.~(\ref{Sinst1QP}), while the LS saturates in the opposite limit. 
This is a consequence of the general property of quantum dynamics near a bifurcation point that can be reduced to fluctuations of a single ``slow'' dynamical variable: since this variable commutes with itself, the only signature of the quantumness is that the noise intensity $T$ is replaced with $(\omega_\tp/2)\coth(\omega_\tp/2T)$.

Like in the classical theory, higher order terms in $\epsilon$ can deform the shape of the potential $U(Q)$.  Unlike the classical theory, they also lead to an additional term in the effective Hamiltonian $K_0$ that is of the form $P^3X$.  The effect of this term is very weak near the bifurcation point and appears only at $O(\epsilon^2)$ corrections to $K_0$.  This originates from the additional nonlinear term $(\bar\phi^\q\phi+\bar\phi\phi^\q)\bar\phi^\q\phi^\q$ in the  Keldysh formulation.
Such a term cannot be reduced to an auxiliary noise field, rendering the theory inequivalent to a Langevin process. As a result, the instanton action is no longer just a height of an effective potential barrier.
This turns out not to have a big effect as the additional factor does not modify the topology of the phase portrait, and thus gives only a weak numerical modification to the phase slip rate.

\section{Conclusion}\label{VI}

Switching between coexisting stable states in Floquet systems is a generic phenomenon. A qualitative aspect of the switching is the possibility, even  for low temperatures, to go over an effective barrier rather than to tunnel. This has no analog in systems in thermal equilibrium. The real-time instanton is a trajectory in the complex phase space of the system, which is the most likely path of  the corresponding quantum fluctuation. Therefore it may not be directly observed.

This paper shows that, albeit indirect, unambiguous demonstration of complex real-time instantons and their features is possible. The suggested way to do it is to use an auxiliary comparatively weak drive, in addition to the strong parametric drive that defines the Floquet dynamics. The auxiliary drive can exponentially strongly modify the switching rate, depending on its frequency. Essentially, it  performs an exponentially-sensitive spectroscopy of the system dynamics as the system moves along the instanton trajectory.

The specific system studied in the paper is a parametric oscillator. In recent years, such an oscillator has been attracting an increasing attention in mesoscopic physics and quantum information. Its stable  states have opposite phases, and switching between them is a phase-slip transition.

We show that the rate of phase-flip transitions is exponentially strongly modified if, in addition to the parametric modulation, the oscillator is driven by an extra field with frequency close to half the modulation frequency. If the oscillator is underdamped, the logarithm of the rate change shows sharp strongly asymmetric peaks. The peaks are almost equidistant, but have different amplitudes. The peaks manifest where the detuning of the extra drive from half the modulation frequency is close to the overtones of the frequency of weakly damped oscillator vibrations about the stable states; such vibrations occur in the rotating frame and modulate the parametrically excited vibrations in the lab frame.

The shapes and amplitudes of the spectral peaks ``probe'' the instanton trajectory. Their amplitudes linearly depend on the amplitude of the extra drive, this is not an Ohmic heating by an extra drive. The peaks' hight  scales as the inverse square root of the decay rate, facilitating the observation of the effect for weak damping. On the other hand, for stronger damping, the spectrum of the response to the extra drive displays a characteristic single peak.

On the technical side, the paper presents a unified approach to the problem of switching based on the Keldysh technique. The formulation naturally connects the quantum and classical analyses. It provides an explicit description of the real-time instantons for weak damping and close to bifurcation points where, unavoidably, the damping becomes effectively strong. It also explicitly describes the effect of the extra drive and makes it possible to study the spectroscopy of real-time instantons in a broad range of parameters, including a broad temperature range.

\section*{Acknowledgments}
FT is supported by the Deutsche Forschungsgemeinschaft (DFG, German Research Foundation) under Germany’s Excellence Strategy Cluster of Excellence Matter and Light for Quantum Computing (ML4Q) EXC 2004/1 390534769 and by the DFG Collaborative Research Center (CRC) 183 Project No. 277101999 - project B02.  FT and AK were supported by the NSF grant DMR-2338819. MID acknowledges partial support from the Moore Foundation Award No. GBMF12214. D.K.J.B. gratefully acknowledges financial support from the Deutsche Forschungsgemeinschaft (DFG, German Research Foundation) through Project-ID 425217212 - SFB 1432.

\appendix
\section{Classical Mechanics of the PDO}\label{appA}
\begin{figure}[t]
    \centering
    \includegraphics{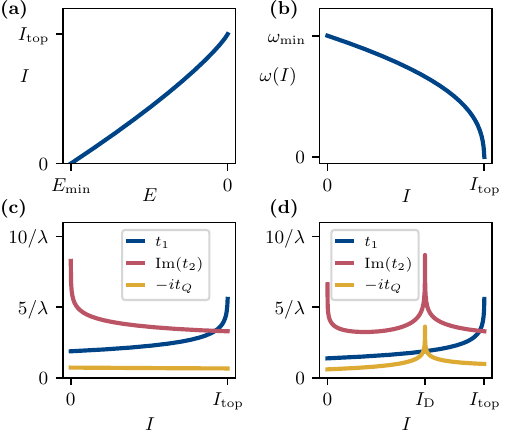}
    \caption{Panel (a) shows the action $I$ as a function of energy $E$. Panel (b) shows the frequency $\omega$ as a function of the action $I$. Panel (c) and (d) show the two periods $t_1$ and $t_2$ and the quasi-period $t_\tQ$ as functions of $I$ for $\Delta/2\lambda=0.3$ and $\Delta/2\lambda=-0.3$ respectively.}
\label{fig:Periods}
\end{figure}

The solutions to the classical motion described by the rotating wave Hamiltonian Eq.~(\ref{HRWA}) were presented originally in \cite{QuantumActivation}; they are reproduced in the notation of this manuscript here.  It is most convenient to define them using the quasi-energy $E$ instead of the action $I$.
Within one well, the quasi-energy is valued in the interval $E\in(E_\tmin,0)$ where $E_\tmin$ is given in the main text, reproduced here for convenience,
\begin{equation}
E_\tmin=-\frac{(2\lambda-\Delta)^2}{g}.
\end{equation}
The action as a function of energy $I(E)$ lacks a simple closed-form expression, but is a monotonic function of $E$ and may be expressed through the integral relation $I(E)=\int_{E_\tmin}^E\dif E'/\omega(E')$; see Fig. \ref{fig:Periods}.
The corresponding limiting forms of the action are $I(E_\tmin)=0$ and $I(E=0)=I_\mathrm{top}$.

The classical solutions $\underline\phi$ have the functional form:
\begin{equation}
\underline\phi(t;E)=\bigg|\frac{E}{g}\bigg|^{1/4}\frac{1}{k_++k_-\mathrm{cn}}\bigg(2\im|gE|^{1/4}\mathrm{dn}-\frac{k_+k_-}{\sqrt{2\lambda}}\mathrm{sn}\bigg),
\end{equation}
where $k_\pm$ are constants and $\mathrm{sn}$, $\mathrm{cn}$, and $\mathrm{dn}$ are the corresponding Jacobi elliptic functions,
\begin{subequations}
\begin{equation}
k_\pm(E)=\sqrt{2\lambda-\Delta\pm\sqrt{g|E|}},
\end{equation}
\begin{equation}
\mathrm{pq}=\mathrm{pq}\Big(\sqrt{8\lambda}|gE|^{1/4}t\Big|m(E)\Big),
\end{equation}
\end{subequations}
with $\mathrm{pq}=\mathrm{sn,cn,dn}$ and the moduli parameter $m$ is:
\begin{equation}
m(E)=\frac{-\big(2\lambda-\Delta-\sqrt{g|E|}\big)\big(2\lambda+\Delta-\sqrt{g|E|}\big)}{8\lambda\sqrt{g|E|}}-\im0^+
\end{equation}

As elliptic functions, $\underline\phi$ have two complex-valued periods, which are given by:
\begin{subequations}\label{Periods}
\begin{equation}
t_1(E)=\sqrt\frac{2}{\lambda\sqrt{g|E|}}\mathrm{K}\big(m(E)\big),
\end{equation}
\begin{equation}
t_2(E)=\im\sqrt\frac{2}{\lambda\sqrt{g|E|}}\mathrm{K}\big(1-m(E)\big).
\end{equation}
\end{subequations}
Here, $\mathrm{K}$ is the complete elliptic integral of the first kind.
Note that for for $\Delta\geq0$ or $\Delta<0$ and $E>E_\tD$, $\mathrm{Re}(t_2)=t_1$ and for $\Delta<0$ and $E<E_\tD$, $\mathrm{Re}(t_2)=0$.
The energy scale $E_\tD$ is a finite energy scale for which $|t_2|\to\infty$ as $E\to E_\tD$, given by:
\begin{equation}\label{app_ED}
E_\tD=-\frac{(2\lambda+\Delta)^2}{g}.
\end{equation}
It is only greater than $E_\tmin$ when $\Delta<0$.  
This defines the corresponding value of the action $I_\tD=I(E_\tD)$. This is also the scale at which the $T=0$ instanton $p_\inst^{(T=0)}$ from Eq.~(\ref{pinstT=0}) diverges, see Fig. \ref{fig:quantumInstanton}.

The periods $t_{1,2}$ are respectively real and imaginary.
As a function of $E$, $t_1$ is a monotonically increasing function and $t_2$ is a monotonically decreasing function when $\Delta>0$ and non-monotonic when $\Delta<0$.  They are plotted for representative values of $\Delta$ in Fig.~\ref{fig:Periods}.

There are three additional complex-valued time arguments for which the classical solutions have special properties.
The function $\underline\phi(t;E)$ describes motion around the well $\phi_0$; motion around the symmetric well $-\phi_0$ is described by $\underline\phi(t+t_\tS;E)$, with $t_\tS$ the complex-valued time:
\begin{equation}\label{ts}
t_\tS(E)=\sqrt\frac{1}{2\lambda\sqrt{g|E|}}\Big(\im \mathrm{K}\big(1-m(E)\big)+\mathrm{K}\big(m(E)\big)\Big).
\end{equation}
The function $\underline\phi$ has a pole at $t_\tP$,
\begin{equation}\label{tS}
t_\tP(E)=\frac{1}{\sqrt{8\lambda}|gE|^{1/4}}\mathrm{arccn}\big(-k_+/k_-\big|m(E)\big),
\end{equation}
which is complex valued and obeys the relations $\mathrm{Re}\big(t_\tP(E)\big)=t_1(E)/2$ and $0<\mathrm{Im}\big(t_\tP(E)\big)<\mathrm{Im}\big(t_2(E)\big)/4$.

The location of the pole is used to define the quasi-period imaginary time value $t_\tQ$ which is needed in the definition of the $T=0$ instanton $p_\inst^{(T=0)}$ from Eq.~(\ref{pinstT=0}),
\begin{equation}\label{tQ}
t_\tQ(E)=t_1(E)+\frac{1}{2}\mathrm{Im}\big(t_2(E)\big)-2t_\tP(E)
\end{equation}
The quasi-period $t_\tQ$ is plotted in Fig.~\ref{fig:Periods}.
Note that $\mathrm{Im}(t_\tQ)\geq0$ and $\mathrm{Re}(t_\tQ)=0$ for all $E$.  From the bounds obeyed by the pole time $t_\tP$ it follows also for all $E$ that $\mathrm{Im}(t_2)/2>|t_\tQ|>0$.
The exact form of the quasi-period condition obeyed by the classical solutions depends on $E$ and $\Delta$.  For $\Delta\geq0$ or $\Delta<0$ and $E>E_\tD$ it is $\underline\phi(t-t_\tQ(E);E)=-\underline{\bar\phi}(t;E)$.  For $\Delta<0$ and $E<E_\tD$ it is instead $\underline\phi(t-t_\tQ(E);E)=-\underline{\bar\phi}(t-t_1(E)/2;E)$.  These relations lead in both cases to the annihilation of the $T=0$ Keldysh Hamiltonian Eq.~(\ref{K0quantum_Integral}) by $p_\inst^{(T=0)}$.

\begin{figure}
    \centering
    \scalebox{.4}{\includegraphics{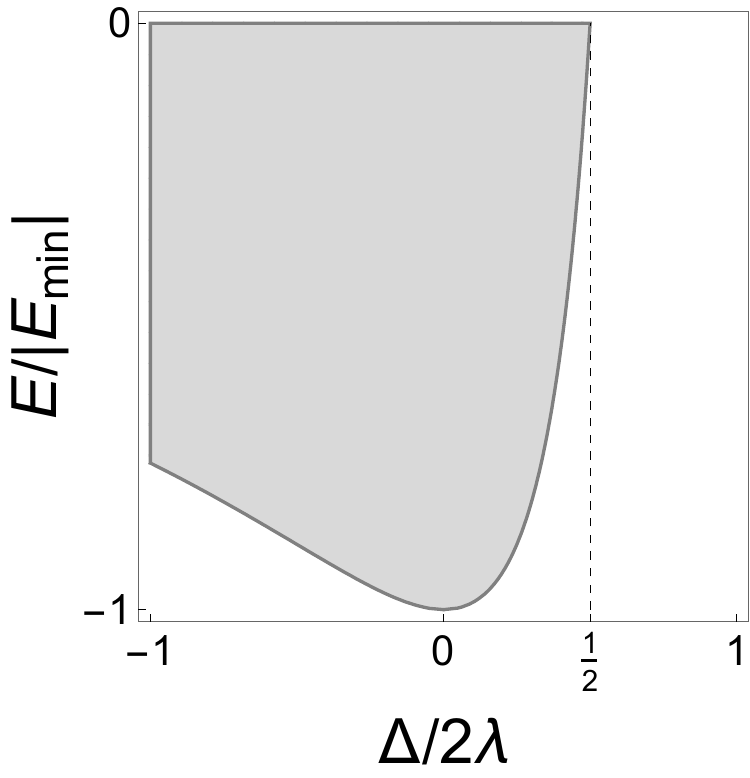}}
    \caption{The shaded region indicates the range of energies for which zero-temperature fragility occurs for a given value of $\Delta/2\lambda$.  The boundary between the shaded and unshaded regions define $E_\tF$.  The dashed line shows the threshold above which there is no fragility.}
    \label{fig:fragilityRegion}
\end{figure}

The Fourier coefficients $c_n(I)$ in the action-angle representation of $\underline\phi$ in Eq.~(\ref{ActionAngle}) are:
\begin{equation}\label{cn}
c_n(I)=\int\limits_0^{2\pi}\frac{\dif\theta}{2\pi}\e^{\im n\theta}\underline\phi(\theta;I)=\im\omega(I)\frac{1}{\sqrt{\lambda g}}\frac{\e^{\im n\omega(I)t_\tP(I)}}{1+\e^{\im n\omega(I)t_\tS(I)}}.
\end{equation}
They have the approximate asymptotic expressions for large $|n|$,
\begin{equation}\label{cnCoefs}
c_n\propto\begin{cases}\e^{-n\omega t_+},&n\gg1\\
\e^{-|n|\omega t_-},& n\ll-1
\end{cases}
\end{equation}
where the positive/negative angles are:
\begin{equation}\label{t><}
t_+=\mathrm{Im}(t_\tP),\qquad t_-=\mathrm{Im}(t_\tS-t_\tP)
\end{equation}
This implies that the left and right radius of convergence of the zero-temperature Keldysh Hamiltonian $K_\ell$ are given by:
\begin{equation}
p_\gtrless(I)=2\omega(I)t_\mp(I).
\end{equation}
The value of the action $I_\tF$ above which zero-temperature fragility occurs is determined by the condition $p_<(I_\tF)=p^{(T=0)}_\inst(I_\tF)$, at which point the $T=0$ instanton is no longer within the reduced radius of convergence of the $T\to0$ Keldysh Hamiltonian.
The corresponding quasi-energy value $E_\tF=H_0(I_\tF)$ may also be defined.  The region of quasi-energies and parameter space where this occurs is depicted in Fig.~\ref{fig:fragilityRegion}.

\section{Limiting Forms of the Classical Solutions}\label{appB}

Near the bottom of the well $E\to E_{\tmin}$ (or equivalently $I\to0$), the problem becomes approximately harmonic.  In this limit the problem can be easily solved by writing $\phi=\phi_0+\delta\phi$ and expanding to quadratic order in $\delta\phi$.  The resulting quadratic Hamiltonian has the same form as the linear rotating wave Hamiltonian Eq.~(\ref{HRWA}),
\begin{equation}\label{LinHam}
H_0\simeq E_{\tmin}+\bar\omega\delta\bar\phi\delta\phi+\frac{\Delta}{2}(\delta\phi^2+\delta\bar\phi^2),
\end{equation}
where $\bar\omega=4\lambda-\Delta$.
The Hamiltonian is diagonal in the coordinates $\zeta=\cosh(\beta)\phi+\sinh(\beta)\bar\phi$, where the Bogoliubov rotation angle is given by:
\begin{equation}
\beta=\frac{1}{2}\arctanh(\Delta/\bar\omega).
\end{equation}
This gives $H_0=E_\tmin+\omega_\tmin\bar\zeta\zeta$, with the the frequency $\omega_\tmin$ matching the limiting form of $\omega(I=0)$.  This, together with its derivative are both finite:
\begin{subequations}\label{OmegaAssymptoticE0}
\begin{equation}
\omega(I=0)\equiv\omega_{\tmin}=2\sqrt{2\lambda(2\lambda-\Delta)},
\end{equation}
\begin{equation}
\partial_I\omega(I=0)=-\frac{g(8\lambda-\Delta)}{4(2\lambda-\Delta)}.
\end{equation}
\end{subequations}
This is shown in the plot of $\omega(I)$ in Fig.~\ref{fig:Periods}.

In this limit, the real period $t_1$ is finite and the imaginary period $t_2$ diverges logarithmically.
The pole time $t_\tP$ also diverges logarithmically but with an overall prefactor of $1/4$ compared to $t_2$.  This leads to a cancellation, rendering the quasi-period time $t_\tQ$ finite (though without a simple closed-form expression),
\begin{subequations}\label{tassymptoticE0}
\begin{equation}
t_1(I=0)=\frac{2\pi}{\omega_\tmin},
\end{equation}
\begin{equation}
|t_2(I)|\underset{I\ll\bar I}{\simeq}\frac{2}{\omega_\tmin}\big|\log(I/\bar I)\big|\toto{I}{0}\infty,
\end{equation}
\begin{equation}
|t_P(I)|\underset{I\ll\bar I}{\simeq}\frac{1}{2\omega_\tmin}\big|\log(I/I_\tP)\big|\toto{I}{0}\infty,
\end{equation}
\begin{equation}
t_\tQ(I)\toto{I}{0}t_\tQ(0)<\infty,
\end{equation}
\end{subequations}
where $\bar I=\omega_\tmin^2/(\lambda g\sinh(\beta)\cosh(\beta)))$ and $I_\tP=\omega_\tmin^2/(\lambda g\sinh^2(\beta))$.

In the harmonic limit the Hamiltonian written in terms of the action adopts the approximate form $H_0\simeq E_\tmin+\omega_\tmin I$.  It follows that the first Fourier coefficients are the largest, with $c_{\pm1}\sim\sqrt I$.
The exact expression in the Harmonic limit is obtained from the diagonalizing Eq.~(\ref{LinHam}),
\begin{equation}
c_1(I\to0)=\cosh(\beta)\sqrt{I},\qquad c_{-1}(I\to0)=\sinh(\beta)\sqrt{I}.
\end{equation}
These may be expressed in terms of the microscopic parameters explicitly using the relations
\begin{subequations}
\begin{equation}
\cosh(\beta)=\sqrt\frac{\bar\omega+\omega_\tmin}{2\omega_\tmin}
\end{equation}
\begin{equation}
\sinh(\beta)=\frac{\Delta}{\sqrt{2\omega_\tmin}\sqrt{\bar\omega+\omega_\tmin}}.
\end{equation}
\end{subequations}

The other Fourier components come with increasing powers of $I$, which can be deduced from the limiting forms of the various periods discussed above.
Substituting the forms of the logarithmic divergences from Eq.~(\ref{tassymptoticE0}) into Eq.~(\ref{cn}) gives, up to an overall complex phase,
\begin{equation}\label{cnI=0}
c_{n}(I)\underset{I\ll I_\gtrless}{\simeq}\frac{\omega_\tmin}{\sqrt{\lambda g}}(I/I_\gtrless)^{|n|/2},
\end{equation}
where $I_>=\bar I^2/I_\tP$ and $I_<=I_\tP$ apply for $n\gtrless0$.

In the opposite limit the classical motion approaches the separatrix near the top of the well.  This corresponds to $E\to0^-$ and $I\to I_\mathrm{top}\equiv I(E\to0^-)$.
In this limit the frequency goes to zero and its derivative diverges,
\begin{subequations}\label{OmegaAssymptotic0}
\begin{equation}
\omega(I)\underset{I\simeq I_\mathrm{top}}{\simeq}\frac{2\pi\sqrt{g\sqrt{E_{\tmin}E_\tD}}}{\big|\log(1-I/I_\mathrm{top})\big|}\toto{I}{I_\mathrm{top}}0,
\end{equation}
\begin{equation}
\partial_I\omega(I)\underset{I\simeq I_\mathrm{top}}{\simeq}\simeq-\frac{E_{\tmin}\big(\omega(I)\big)^2}{E(I)\big|\log(1-I/I_\mathrm{top})\big|}\toto{I}{I_\mathrm{top}}-\infty.
\end{equation}
\end{subequations}
The physical period $t_1$ diverges as the separatrix is approached.  The imaginary period $t_2$ and the quasi-period $t_\tQ$ both have finite limits.
\begin{subequations}\label{tAssymptotic0}
\begin{equation}
t_1(I)\underset{E\simeq0^-}{\simeq}\frac{1}{\sqrt{g\sqrt{E_{\tmin}E_\tD}}}\big|\log(1-I/I_\mathrm{top})\big|\toto{I}{I_\mathrm{top}}\infty,
\end{equation}
\begin{equation}
\mathrm{Im}\big(t_2(I_\mathrm{top})\big)=\frac{2\pi}{\sqrt{g\sqrt{E_{\tmin}E_\tD}}},
\end{equation}
\begin{equation}
t_\tQ(I)\toto{I}{I_\mathrm{top}}t_\tQ(I_\mathrm{top})<\infty.
\end{equation}
\end{subequations}
Similar to $t_\tQ$, the complex symmetric time $t_\tS$ and the pole time $t_\tP$, and consequently also the limiting times in the Fourier coefficients $t_\gtrless$, have finite limits as $I\to I_\mathrm{top}$.
There are similar logarithmic singularities in both $t_2$ and $t_\tQ$ near $I_\tD$ when $\Delta<0$, see Fig. \ref{Periods}.

\section{Prefactors of the LS Harmonics}\label{appC}
The form of the Fourier coefficients in the harmonic limit discussed in Appendix \ref{appB} can be used to derive the limiting forms of various functions used in the main text.  This enables the extraction of the leading order form of the LS resonances, both classical, Eq.s~(\ref{LSharmonicResonanceClassical}) and (\ref{LSharmonicResonanceClassical-n}), and quantum, Eq.~(\ref{LSharmonicResonanceQuantum-all}).

From the definition Eq.~(\ref{eq:sum-rules}), the classical noise kernel $\Gamma(I)$ is seen to acquire the form:
\begin{equation}
\Gamma(I\to0)\simeq|c_1|^2+|c_{-1}|^2=\frac{\bar\omega}{\omega_\tmin}I,
\end{equation}
where $\bar\omega=4\lambda-\Delta$ as defined above.
The fixed point $p_*=p_\inst(I=0)$ can be obtained from solving $\partial_IK_0(I=0)=0$ for $p$.  For the classical theory this gives $p_*=\omega_\tp/(T\partial_I\Gamma(0))$, which upon inputting the above expression becomes:
\begin{equation}
p_*=\frac{\omega_\tp}{T}\frac{\omega_\tmin}{\bar\omega}.
\end{equation}
The numerical prefactor $a_n$ in front of the leading-order term in $\chi_n$ from Eq.~(\ref{LSharmonicResonanceClassical-n}) may be obtained from the leading-order term in $\chi_n$ for $\nu$ near $n\omega_\tmin$, which gives $a_n=(c_n/I^{|n|/2})|_{I=0}\sqrt{|n|}Tp_*/\omega_\tp$.  Using the above expression and Eq.~(\ref{cnI=0}), one obtains:
\begin{equation}
a_n=\sqrt{|n|}\frac{\omega_\tmin}{\bar\omega}\bigg(\frac{\lambda g}{\omega_\tmin^2}\bigg)^\frac{|n|-1}{2}
\begin{cases}
\cosh^{|n|}(\beta),&n>0\\
\sinh^{|n|}(\beta),&n<0
\end{cases}
\end{equation}
For $n=1$ this reproduces Eq.~(\ref{a1}).

In the quantum theory, the single-step transition rates from Eq.~(\ref{Wn}) are
\begin{subequations}\label{WsmallI}
\begin{equation}
W_1(I\to0)\simeq\big(\gamma_\ell\cosh^2(\beta)+\gamma_\tg\sinh^2(\beta)\big)I,
\end{equation}
\begin{equation}
W_{-1}(I\to0)\simeq\big(\gamma_\ell\sinh^2(\beta)+\gamma_\tg\cosh^2(\beta)\big)I.
\end{equation}
\end{subequations}
From this, an expression for the fixed point $p_*$ is given by $p_*=\log(\partial_IW_1(0)/\partial_IW_{-1}(0))$, which may be evaluated to:
\begin{equation}\label{p*quantum}
p_*=\log\bigg(\frac{\bar\omega(\gamma_\ell+\gamma_\tg)+\omega_\tmin(\gamma_\ell-\gamma_\tg)}{\bar\omega(\gamma_\ell+\gamma_\tg)-\omega_\tmin(\gamma_\ell-\gamma_\tg)}\bigg).
\end{equation}

To determine the prefactor $A_n$, an expression for the sum in the denominator of Eq.~(\ref{LSharmonicQuantum}) is also needed.  In the harmonic limit it adopts the simple form $-\partial_I(W_1\e^{-p_*}-W_{-1}\e^{p_*})|_{I=0}$, which upon insertion of the above formulas gives exactly 1.  In other words, $\partial_tI_\inst\simeq2\kappa I_\inst$ at early times in the quantum problem even for finite temperatures.  This is true despite the lack of detailed balance.  This holds however only in the limit of early times and is violated along the remainder of the instanton trajectory for $T>0$.

This implies the prefactor for the limiting for of the quantum LS has the form $A_n=(c_n/I^{|n|/2})|_{I=0}2\sinh(np_*/2)|n|^{-1/2}$.
Using the above expressions, this leads to:
\begin{multline}
A_n=\frac{\big((2n_\tB+1)\bar\omega+\omega_\tmin\big)^{|n|}-\big((2n_\tB+1)\bar\omega-\omega_\tmin\big)^{|n|}}{\sqrt{|n|}\big((2n_\tB+1)^2\bar\omega^2-\omega_\tmin^2\big)^{|n|/2}}\\
\times\bigg(\frac{\lambda g}{\omega_\tmin^2}\bigg)^\frac{|n|-1}{2}
\begin{cases}
\cosh^n(\beta),&n>0\\
\sinh^{|n|}(\beta),&n<0
\end{cases}
\end{multline}
where the definitions of the dissipation rates in terms of bath occupation number $\gamma_\ell=2\kappa(n_\tB+1)$ and $\gamma_\tg=2\kappa n_\tB$ have been employed.
Eq.~(\ref{A1}) is retrieved by putting $n=1$.

\section{Table of correspondence}
\label{sec:table}

The dynamics of quantum parametric oscillators have been described using different notations. Here we provide the correspondence between the notations in the present paper and in the previous work on quantum activation, in particular \cite{QuantumActivation,QuantumSymmetryLiftingZeroT}. In Reference \cite{QuantumActivation} there was no extra field with frequency $\omega_\td\approx \omega_\tp/2$, whereas in Reference \cite{QuantumSymmetryLiftingZeroT} the extra field was just at  half the frequency of parametric modulation, therefore no parameters of the extra field are included in the table \ref{table:comparison}.  The dynamical variables in the rotating frame in the present paper are $\phi$ and $\bar\phi$, whereas in \cite{QuantumActivation,QuantumSymmetryLiftingZeroT} there were used the scaled coordinate $Q$ and momentum $P$. We give the expressions for the dynamical variables in terms of the ladder operators $\hat a, \hat a^\dagger$. The results of \cite{QuantumSymmetryLiftingZeroT} are given for the oscillator mass in the laboratory frame $m_0 = 1$.  The notation for the coefficient of linear (viscous) friction in the laboratory frame in \cite{QuantumActivation,QuantumSymmetryLiftingZeroT} is $\Gamma$. We give the expressions for the parameters in \cite{QuantumActivation,QuantumSymmetryLiftingZeroT} for the Duffing (Kerr) nonlinearity $\gamma >0$.

\begin{widetext}
	\begin{table*}[t]
		\caption{The relation to the parameters used in some of the previous work on quantum activation}
		\label{tab:example}
		\begin{ruledtabular}
			\renewcommand{\arraystretch}{1.5}
			\begin{tabular}{lccc}
				&Present paper &Reference \cite{QuantumActivation} & Reference \cite{QuantumSymmetryLiftingZeroT}  \\
				\hline
				Modulation frequency & $2\omega_\mathrm{p}$& $\omega_F$ &$ \omega_p$\\
				Modulation amplitude and scaled amplitude&$A_\tp,  \lambda = A_\mathrm{p}/8\omega_\mathrm{p}$ & $F$ & $F_p$ \\

				Amplitude and scaled amplitude of the weak drive
				& $A_\mathrm{d}, \;\alpha = A_\mathrm{d}/(\sqrt{8\omega_\mathrm{p}})$
				& 
				& $A_d, \alpha_d = A_d \sqrt{6\gamma/F_p^3}$\\
				
				Duffing (Kerr) nonlinearity parameter& $G, \;  g = 3G/2\omega_\mathrm{p}^2$ &$\gamma$ & $\gamma$\\
				
				Reduced Planck constant & $\hbar = 1$ & $\lambda = 3\gamma\hbar/F\omega_F$ & $\lambda = 3\gamma \hbar/F_p \omega_p$\\
				RWA Hamiltonian in the absence of dissipation&$ H_0$ & $\hat  g = (6\gamma/F^2) H_0 $ & $\hat g = (6\gamma/F_p^2)H_0$\\

				Variables that correspond to the ladder operator $a$ & $\phi $ & $(P-iQ)/(2\lambda)^{1/2} $ & $ (P-iQ)/(2\lambda)^{1/2} $ \\
				Fourier expansion of the variables & $\sum c_n(I)e^{-in\theta}$ &
				$\sum a_m(g)e^{im\phi}$ &$\sum a_m(g)e^{im\phi}$ \\
				
				Detuning $-\Delta$  and scaled detuning $\mu$ of half the modulation &$-\Delta = \frac{\omega_\mathrm{p}^2 - \omega_0^2}{2\omega_\mathrm{p}} $  & $\mu = \frac{\omega_F(\omega_F-2\omega_0)}{F} $ &$\mu = \frac{\omega_p (\omega_p - 2\omega_0)}{F_p}$\\
				frequency from  the oscillator eigenfrequency $\omega_0$ & &   &\\
				Decay rate and scaled decay rate & $\kappa$ & $\eta = 2\omega_F\Gamma/F $ & $ \kappa = 2\omega_p \Gamma/F_p$\\
				Thermal occupation number of the oscillator & $n_\tB$ & $\bar n $ & $\bar n$\\
				Exponent of the quantum activation phase-slip rate $W_\mathrm{ps}$ & $ iS_\mathrm{inst}$ &$ R_A/\lambda$& $R_A/\lambda$

			\end{tabular}
		\end{ruledtabular}
		\label{table:comparison}
	\end{table*}
\end{widetext}

\section{Relation to the Hamiltonian formulation}
\label{sec:Hamiltonian}

The approach based on the Keldysh technique has much in common with the analysis of the switching rates between the states of forced vibrations, which is based on the Hamiltonian formulation \cite{Dykman1988a}. The latter analysis is also related to the approach to the rare events in classical systems described by the Fokker-Planck equation, cf.~\cite{Graham1973,Graham1983a,Graham1985c,Chinarov1993,Maier1993}. A qualitative difference between the classical and quantum formulations  is that, in the quantum problem, the extreme trajectory of the appropriate Hamiltonian, which describes the optimal path followed by the system in switching, i.e., the real-time instanton, is complex. Therefore it is not directly observable, and the logarithmic susceptibility is a major tool for studying  its features.

To reveal the similarity of the Hamiltonian-based and Keldysh-technique based approaches, we will re-write the Hamiltonian approach \cite{Dykman1988a} directly in the action-angle variables. There is no difference, in general terms, between the formulations of the problem  of a resonantly driven oscillator considered in \cite{Dykman1988a} and the problem of a parametrically modulated oscillator considered here, although the solutions are very different. This is rooted in the different symmetries: a resonantly driven oscillator does not have parity. Since the logarithmic susceptibility was not discussed in \cite{Dykman1988a}, we will begin with the case where the only time-dependent drive is parametric modulation.   

The starting point of the Hamiltonian formulation is to write the Lindblad equation for the density matrix $\rho\equiv \rho(Q_1,Q_2)$  of a modulated oscillator in the coordinate representation. In the variables and the parameters of the 3rd column of Table~\ref{table:comparison} the equation reads
\begin{align}
	\label{eq:H_Lindblad} 
	&
	\partial_\tau \rho = -i\lambda^{-1}\cH(Q_1, Q_2,-i\lambda\partial_{Q_1}, -i\lambda\partial_{Q_2})\rho,\nonumber\\
	&\cH(Q_1,Q_2,P_1,P_2) = \hat g(Q_1,P_1) - \hat g(Q_2,P_2) \nonumber\\
	&+ \kappa\hat\cD(Q_1,Q_2,P_1,P_2).
\end{align}
Here, $\tau = tF_p/2\omega_p$ is the dimensionless time, $\lambda$ and $\kappa$ are the dimensionless reduced Planck constant and the decay rate, 
\begin{align}
	\label{eq:g_Hamiltonian}
	\hat g =  \frac{1}{4} (Q^2+P^2-\mu)^2 + \frac{1}{2} (P^2 - Q^2) -\frac{\mu^2}{4}
\end{align}
is the scaled RWA Hamiltonian of the oscillator in the absence of dissipation, and
\begin{align}
	\label{eq:dissipation}
	&\hat \cD=-( P_1Q_2+ P_2 Q_1)\nonumber\\ 
	&- \frac{1}{2}i(2\bar n+1)[(Q_1 -Q_2)^2 + (P_1 + P_2)^2] + i\lambda
\end{align}
is the operator that describes relaxation (linear damping). It corresponds to the operator $\sum \gamma_v D_v$ in Eq.~\eqref{eq_ActionS_full}. 

We assume that $\lambda\ll 1$. Then each ``well'' of  the Hamiltonian $g(Q,P)\propto H_0$ has many quasienergy levels (function $H_0$ is plotted in Fig.~\ref{fig:H0Surface}; as seen from Table~\ref{table:comparison}, the coordinate $Q$ is $\propto \mathrm{Im}\,\phi$). For small $\lambda$, there is a separation of time scales: the dwell time inside a well, which is the reciprocal of the phase-slip rate $W_\mathrm{ps}$, is much longer than the relaxation time inside the well, which is $\sim\kappa^{-1}$.  Then the distribution over the intrawell states is quasistationary. To find it, for a given well, and the rate of switching from this well, one can set $\partial_\tau \rho = 0$. In the classical theory of switching, this approximation  can be traced back to Kramers \cite{Kramers1940}. 

The tail of the intrawell distribution and the rate $W_\mathrm{ps}$ can be sought in the eikonal approximation by setting 
\begin{align}
	\label{eq:eikonal_form}
	\rho(Q_1,Q_2) = \exp[iS(Q_1,Q_2)/\lambda].
\end{align}
To logarithmic accuracy, function $S$ can be associated with the action of a classical system with complex coordinates $Q_1,Q_2$ \cite{Dykman1988a}. The equation of motion of this system is the Hamilton-Jacobi equation
\begin{align}
	\label{eq:Hamilton_Jacobi}
	\cH(Q_1,Q_2, \partial_{Q_1}S, \partial_{Q_2}S) = 0,
\end{align}
where $\cH$ is a function, not an operator; the term $i\lambda$ in $\cD$ should be dropped in this approximation. The condition $\cH=0$ is the condition of the stationarity of the density matrix. 

For the considered occupied well, the function $\rho(Q_1,Q_2)$ is Gaussian near the classical stable state of the oscillator in this well $Q_1=Q_2 = Q\st$, with $\partial_{Q_1}S = -\partial_{Q_2}S =P\st$. The values of $Q\st, P\st$  are given by the equations $\partial_{Q_j}\cH= \partial_{P_j}\cH = 0$ with $j=1,2$. Respectively, $S$ is quadratic in $Q_1-Q\st, Q_2 - Q\st$. It can be easily found from Eq.~\eqref{eq:Hamilton_Jacobi}. One can see from the normalization condition $\int dQ_1 dQ_2\, \rho(Q_1,Q_2) =1$ that $S(Q\st,Q\st)=0$, to the leading order in $\lambda$. 

The value of $S$ further away from the  stable state can be found by solving the Hamiltonian equations of motion 
\begin{align}
	\label{eq:eom_Q_P}
	\dot Q_j = \partial_{P_j}\cH, \quad\dot P_j = -\partial_{Q_j}\cH,\quad j=1,2.
\end{align}
This is somewhat different from the Keldysh technique where the ``left'' and ``right''  variables $\phi^+$ and $\phi^-$ evolve in the opposite directions of time. The initial conditions for the equations of motion are $Q_1(\tau), Q_2(\tau) \to Q\st$ and $P_1(\tau), -P_2(\tau) \to P\st$ for $\tau \to -\infty$. In Eq.~(\ref{eq:eom_Q_P}) and below in this section the over-dot means differentiation over $\tau$, i.e., $\dot X \equiv dX/d\tau$.

As in the main text, and as it was done in \cite{Dykman1988a}, we will consider the case of the small damping parameter $\kappa$, although the general formulation based on Eq.~\eqref{eq:eom_Q_P} applies for an arbitrary $\kappa$ (some numerical results on solving the boundary-value problem (\ref{eq:eom_Q_P}) for a resonantly driven oscillator have been discussed in \cite{Lee2025}). We will change from $(Q_j,P_j)$ to the action-angle variables  $(I_j,\varphi_j)$ of dissipation-free motion. The replacement $(Q,P)\to (I,\varphi)$ is described by the standard canonical transformation for the Hamiltonian function $g(Q,P)$. The variables $Q,P$ are periodic in $\varphi$,  
\[Q(I,\varphi) = \sum_mQ_m(I)e^{im\varphi}, \quad P(I,\varphi) = \sum_m P_m(I)e^{im\varphi},\]
and we choose $Q(I,\varphi)=Q(I, -\varphi), \; P(I,\varphi) = - P(I, -\varphi)$.

In the action-angle variables, the equations of motion \eqref{eq:eom_Q_P} read
\begin{align}
	\label{eq:change_of_variables}
	\dot I_j = -\kappa \partial_{\varphi_j}\cD, \quad \dot \varphi_j = (-1)^{3-j}\omega(I_j) +\kappa\partial_{I_j}\cD
\end{align}
with $\omega(I) = \partial g/\partial I$ (in \cite{Dykman1988a} there was used a formulation in terms of quasienergies and phases). Importantly, the variables $I_j,\varphi_j$ on the trajectories \eqref{eq:change_of_variables} are complex. We further change to 
\[ I_j=I+ (-1)^{3-j}I^q/2,\quad  \varphi_j = (-1)^{3-j}\theta + \theta^q/2.\]

From the condition $\cH=0$ it follows that $g(I+I^q/2) - g(I-I^q/2) = -\kappa \cD$. Therefore $I^q \approx -[\kappa/\omega(I)]\cD$ is small and can be disregarded when calculating $\cD$, i.e., $\cD$ is a function of $I \approx I_1 \approx I_2$.  
Then, from Eq.~\eqref{eq:change_of_variables},  $\dot\theta \approx \omega(I)$, whereas $|\dot\theta^q |\propto \kappa \ll \omega(I)$. For a given $I$, the function $\cD$ is periodic in $\varphi_{1,2}$, and therefore
\begin{align*}
	& \cD(Q_1,Q_2,P_1,P_2) = \sum_{m_1,m_2}\cD_{m_1m_2}(I)e^{i(m_1 + m_2)\theta^q/2}\\
	&\times \exp[i(m_1 -m_2)\theta].
\end{align*}
The terms $\propto \exp[i(m_1 - m_2)\theta]$  with $m_1 \neq m_2$ are fast oscillating at frequency $\omega(I)$. Therefore, from Eq.~\eqref{eq:change_of_variables}, $\dot I^q =-\kappa\partial_\theta\cD$ is a sum of fast oscillating terms.  Given that  $I^q\to 0$ for $\tau\to-\infty$, we see that $I^q$ is not just small, but is also fast oscillating, with no components that smoothly depend on time. Then, noting that the condition $\cH=0$ implies, in particular, that there are no fast-oscillating terms in $\cH$, we obtain 
\begin{align}
	\label{eq:balance_eq_implicit}
	\sum_m \cD_{mm}(I)\exp(im\theta^q) = 0.
\end{align}
Setting, similar to the main text, $\theta_q = -ip$ and using the explicit form of $\cD(Q_1,Q_2,P_1,P_2)$, we can write this equation in the form 
\begin{align}
	\label{eq:balance_equation}
	&\sum_m\left[ \bar n |a_m|^2 (1-e^{-mp} )\right.\nonumber\\
	&\left. + (\bar n +1) |a_{-m}|^2(1-e^{-mp}\right] = 0,
\end{align}
where
\[a_m \equiv(P_m-iQ_m)/\sqrt{2\lambda} = -a_m^* .\]
The Fourier component  $a_m \equiv a_m(I)$ is the matrix element of the ladder operator $a$ on the semiclassical Floquet intrawell wave functions $\ket{n}$ of the Hamiltonian $\hat g$, i.e., $a_m(I_n) = \braket{n+m|a|n}$ \cite{QuantumActivation}.

Equation \eqref{eq:balance_equation} determines the function $p\equiv p(I)$. It  coincides with Eq.~(\ref{K0quantum}) of the main text once one takes into account that $a_m(I) = c_{-m}(I)$. In fact, this equation coincides with the balance equation for the populations $\rho_{nn}\equiv \rho(I_n)$ of the intrawell Floquet states \cite{Dykman1988a,QuantumActivation}. The function $p\equiv p(I)$ determines the change of the populations with the varying state number in the continuous limit, $p(I) = -\lambda d\log \rho(I)/dI$. This is seen from Eq.~(\ref{eq:eikonal_form}) for the density matrix as a function of $I$ by noting that, for $\cH=0$, the action $S$ is
\begin{align}
	\label{eq:action_vs_action}
	&S=-\sum_j\int_0^I\varphi_j\, dI_j = -\int \theta \, dI^q  + i\int_0^I p(I')\, dI',\nonumber\\
	&\rho(I) =C \exp\left[-\int_0^I p( I')\,dI'/\lambda\right],
\end{align}
where we have disregarded small corrections from the integral over $dI^q = \dot I^q dt$ (we recall that $I^q$ is small and fast oscillating in time); we have set $S=0$ at the stable state; the prefactor $C$ is a smooth function of $I$.

The rate of phase-slips via quantum activation $W_\mathrm{ps}$ is determined by the population near the top of the barrier that separates the stable vibrational states of the oscillator. This is a saddle point with  $Q_1 = Q_2 = Q_\mathrm{u}$, $P_1 = -P_2 = P_\mathrm{u}$ and, respectively, $I=I_\mathrm{u}$.  For the considered weak damping
\begin{align}
	\label{eq:activation_energy}
	&W_\mathrm{ps} = \mathrm{const}\times\exp(-R_A/\lambda), \nonumber\\
	&R_A = \mathrm{Im}\,S(I_\mathrm{u}) =\int_0^{I_\mathrm{u}}p(I)\, dI
\end{align}
This expression coincides with the result of the main text.

\subsection{Logarithmic susceptibility}
\label{subsec:LS}

In the absence of an extra drive the trajectory followed in escape is described by $Q_j(\tau)\equiv Q_j(I_j(\tau),\varphi_j(\tau)$,  $P_j(\tau)\equiv P_j(I_j(\tau),\varphi_j(\tau)$ ($j=1,2$), where, for weak damping, $I_{1,2}(\tau)$ and $\varphi_{1,2}(\tau)$ are determined by Eq.~(\ref{eq:change_of_variables}) with
\begin{align}
    \label{eq:eom_H}
&\dot I = 2\kappa\lambda \sum_n |a_n|^2 n \left[(\bar n +1)e^{pn} - \bar n e^{-pn}\right],\nonumber\\
&\dot\theta = \omega(I).
\end{align}
Here $\theta^q = -ip\equiv -ip(I)$ is given by Eq.~\eqref{eq:balance_equation} and $a_n\equiv a_n(I) = (P_n - iQ_n)/\sqrt{2\lambda}$. The escape trajectory is a complex real-time instanton. It starts from the dynamically stable classical oscillator state $Q_1=Q_2=Q\st$, $ P_1 = -P_2 = P\st$ and goes to the saddle point (unstable stationary state) $Q_1 = Q_2 = Q_\mathrm{u}$, $P_1 = -P_2 = P_\mathrm{u}$. 

In the notations of Ref.~\cite{QuantumSymmetryLifting,QuantumSymmetryLiftingZeroT} the perturbation from an extra drive at frequency $\omega_d$ close to half the modulation frequency  is described by adding to $ \hat g $ a term $\hat g\1$,
\begin{align}
	\label{eq:perturbation}
	&\hat g\1(Q,P;\tau) = -\alpha_d\left[P\cos(\nu_d\tau + \varphi_d)\right.\nonumber\\
    &\left.+ Q\sin(\nu_d\tau + \varphi_d)\right] 
\end{align}
where $\nu_d = 2\omega_p[\omega_d- (\omega_p/2)]/F_p$, in the notations of the right column of Table~\ref{table:comparison}; $\alpha_d$ is also defined in the same column (note that $\phi_d$ is shifted by $\pi$ compared to Eq.~(\ref{HamPDO}). The corresponding term in the Hamiltonian $\cH$ is
\[\cH\1(\tau) = g\1(Q_1,P_1;\tau) - g\1(Q_2,-P_2;\tau).\]
In terms of the Fourier components $a_m(I)$ and the variables $\theta$ and $p=i\theta^q$ it has the form
\begin{align}
    \label{eq:H1}
    \cH\1(\tau) =& 2\alpha_d(\lambda/2)^{1/2}\sum_m a_{-m}e^{i\nu_d\tau + i\phi_d - im\theta} \nonumber\\
   &\times \sinh (mp/2) -\mathrm{c.c.}
\end{align}

To the first order in $\alpha_d$ the correction to the action due to the perturbation is 
\[S\1 = -\int_{-\infty}^\infty d\tau \,\cH\1(\tau|\tau_0),\]
where $\cH\1(\tau|\tau_0)$ is the Hamiltonian $\cH\1(\tau)$ calculated for the unperturbed trajectory $Q_{1,2}(\tau - \tau_0)$, $P_{1,2}(\tau - \tau_0)$, and one has to take the extremum of $S\1$ over $\tau_0$. This maps the calculation on the calculation of the logarithmic susceptibility in the main text.

\bibliography{Biblio}

\end{document}